\documentclass[twocolumn,trackchanges]{aastex6}
 
\usepackage{hyperref}
\usepackage{graphicx}
\usepackage{subfigure}

\shorttitle{Determination of pulsation periods and other parameters of 2875 stars classified as MIRA in ASAS}
\shortauthors{Vogt et al.}

\begin{document}


\title{Determination of pulsation periods and other parameters of 2875 stars\\ classified as MIRA in the All Sky Automated Survey (ASAS)}


\author{N. Vogt$^{1}$, A. Contreras-Quijada$^{1}$, I. Fuentes-Morales$^{1}$, S. Vogt-Geisse$^{2}$, C. Arcos$^{1}$, C. Abarca$^{1}$,\\ C. Agurto-Gangas$^{1,3}$, M. Caviedes$^{1}$, H. DaSilva$^{1}$, J. Flores$^{1,4}$, V. Gotta$^{1}$,\\ F. Pe\~{n}aloza$^{1}$, K. Rojas$^{1}$ \& J. I. Villase\~{n}or$^{1,5}$}

\affil{$^{1}$Instituto de F\'isica y Astronom\'ia, Facultad de Ciencias, Universidad de Valpara\'iso, Chile\\$^{2}$Departamento de Fisicoqu\'imica, Facultad de Ciencias Qu\'imicas, Universidad de Concepci\'on, Chile\\$^{3}$Max-Planck-Institut f\"ur extraterrestrische Physik (MPE), D-85748 Garching, Germany\\
$^{4}$Departamento de F\'isica, Universidad T\'ecnica Federico Santa Mar\'ia, Valpara\'iso, Chile\\
$^{5}$Departamento de F\'isica y Astronom\'ia, Universidad de La Serena, La Serena, Chile}



\begin{abstract}
\noindent
We have developed an interactive PYTHON code and derived crucial ephemeris data of 99.4\% of all stars classified as 'Mira' in the ASAS data base, referring to pulsation periods, mean maximum magnitudes and, whenever possible, the amplitudes among others. We present a statistical comparison between our results and those given by the AAVSO International Variable Star Index (VSX), as well as those determined with the machine learning automatic procedure of \citealt{r12}. Our periods are in good agreement with those of the VSX in more than 95\% of the stars. However, when comparing our periods with those of Richards et al, the coincidence rate is only 76\% and most of the remaining cases refer to aliases. We conclude that automatic codes require still more refinements in order to provide reliable period values. Period distributions of the target stars show three local maxima around 215, 275 and 330 d, apparently of universal validity; their relative strength seems to depend on galactic longitude. Our visual amplitude distribution turns out to be bimodal, however 1/3 of the targets have rather small amplitudes (A $<$ 2.5$^{m}$) and could refer to semi-regular variables (SR). We estimate that about 20\% of our targets belong to the SR class. We also provide a list of 63 candidates for period variations and a sample of 35 multiperiodic stars which seem to confirm the universal validity of typical sequences in the double period and in the Petersen diagrams.\\
\end{abstract}


\section{Introduction}

Mira variables are evolved pulsating late-type star at the last evolutionary state before the planetary nebula phase. They belong to the Asymptotic Giant Branch (AGB) and therefore they enrich the interstellar material with gas and dust that has undergone nuclear processing. This interaction with their surroundings, due to their strong winds, plays an important role in the Milky Way recycling processes. The mass loss of Mira stars could be accompanied by period changes, which, in some cases, exceed 30\% of the original period value within a few decades (\citealt{s13}). This period variability is normally attributed to He shell flash pulses, but alternative explanations could also be considered (\citealt{t05}, and references therein).  An analysis of light curves and the determination of pulsation periods and other observational parameters provides vital information about the interior processes in stars and their evolution.  However, systematic studies of observational parameters of Mira stars are still scarce.

One of the possibilities to improve this situation is offered by the ASAS database (All-Sky Automated Survey, \citealt{p02}) which is the outcome of a sky patrol program with CCD photometry, conducted at the Las Campanas Observatory, Chile, between 2000 and 2009 (-90$^{\circ}$ $\leq$ $\delta$ $\leq$ +28$^{\circ}$). The visual limiting magnitude is about 14.5$^{m}$ and a total of up to 500 measurements per star are available. An automatic procedure identified 50122 ASAS variable stars and classified part of them according to their type of variability. A total of 2895 variables are listed as Mira stars in this database. However, their pulsation periods, as suggested by ASAS, are not at all reliable, and therefore they cannot be used for a statistical study of Mira properties. The aim of this paper is to re-analyze the light curves of all stars classified as Mira in ASAS, in order to offer to the community reliable periods, amplitudes and other observational parameters (including mean errors), in a way that they can easily be combined with future studies of these stars. For this purpose we used a simple interactive method which is not fully automatic but requires human control and interaction, as explained in Section 2. Section 3 corresponds to our data presentation, while in Section 4 we give a statistical analysis of our results, with emphasis in a comparison to the International Variable Star Index (VSX)  of  the American Association of Variable Star Observes (AAVSO, \citealt{w06}),  as well as to the fully automated machine learning method of \citet{r12}. We also estimate the SR fraction in our sample, select candidates for period changes and give some examples of multiperiodicity. The final Section 5 contains our conclusions.

\section{A simple method determining ephemeris data}

\begin{figure}[hbtp]
\centering
\begin{tabular}{ll}
\includegraphics[scale=0.4]{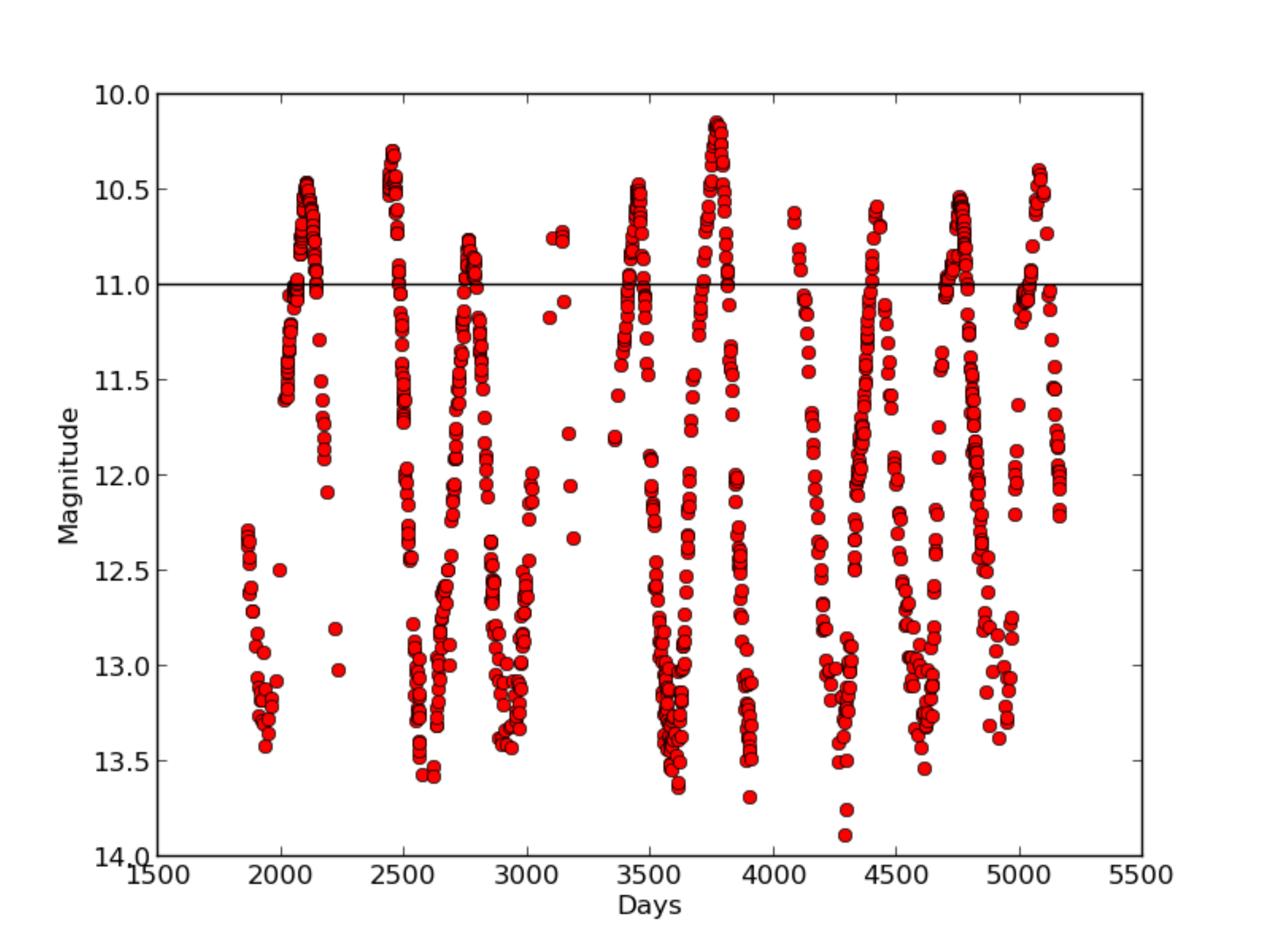} \\
\includegraphics[scale=0.4]{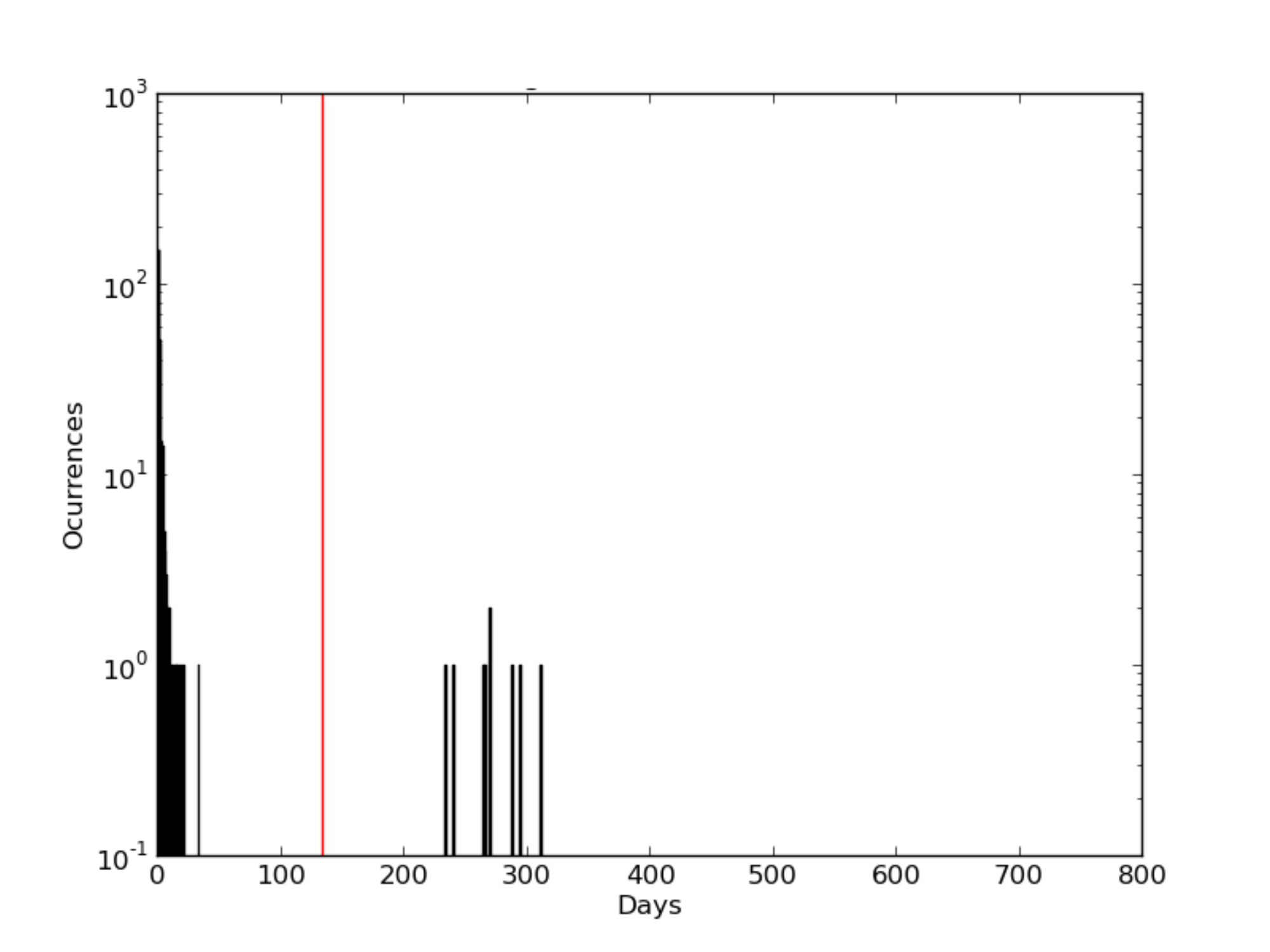}
\end{tabular}
\caption{The first two steps of our method. \textbf{Upper panel}: The light curve of star ASAS 174436-8629.2 (Z Oct). The horizontal cut line serves to select data points around each maximum. Only values brighter than this limit are used in the period determination process. The time is given in HJD - 2450000. \textbf{Lower panel}: The histogram shows the distribution of the time differences between subsequent data points in the light curve; the red line refers to the limit for the cycle counting process (details see text).\label{Fig1}}
\end{figure}

\begin{figure}[hbtp]
\centering
\begin{tabular}{lll}
\includegraphics[scale=0.4]{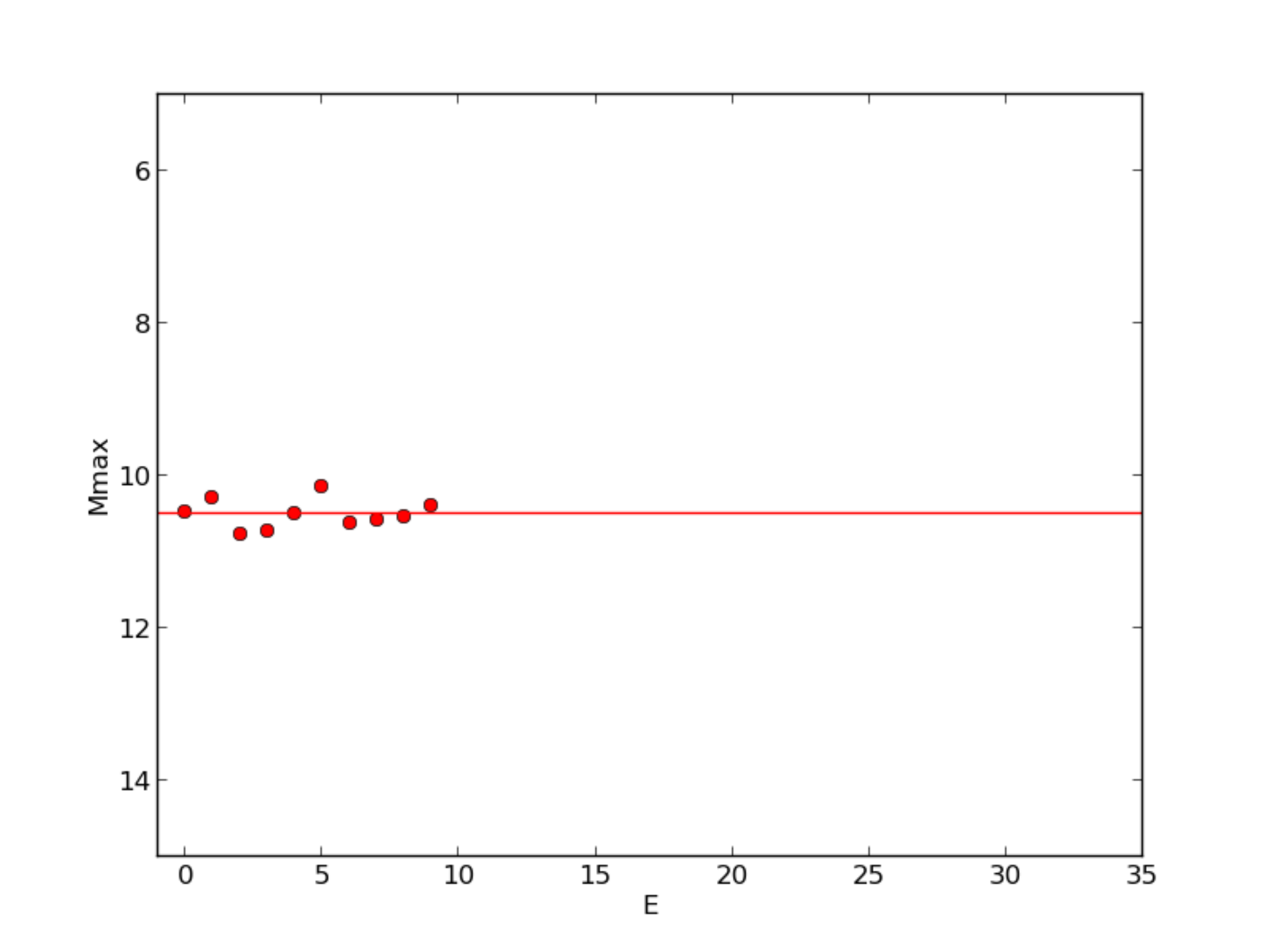} \\
\includegraphics[scale=0.4]{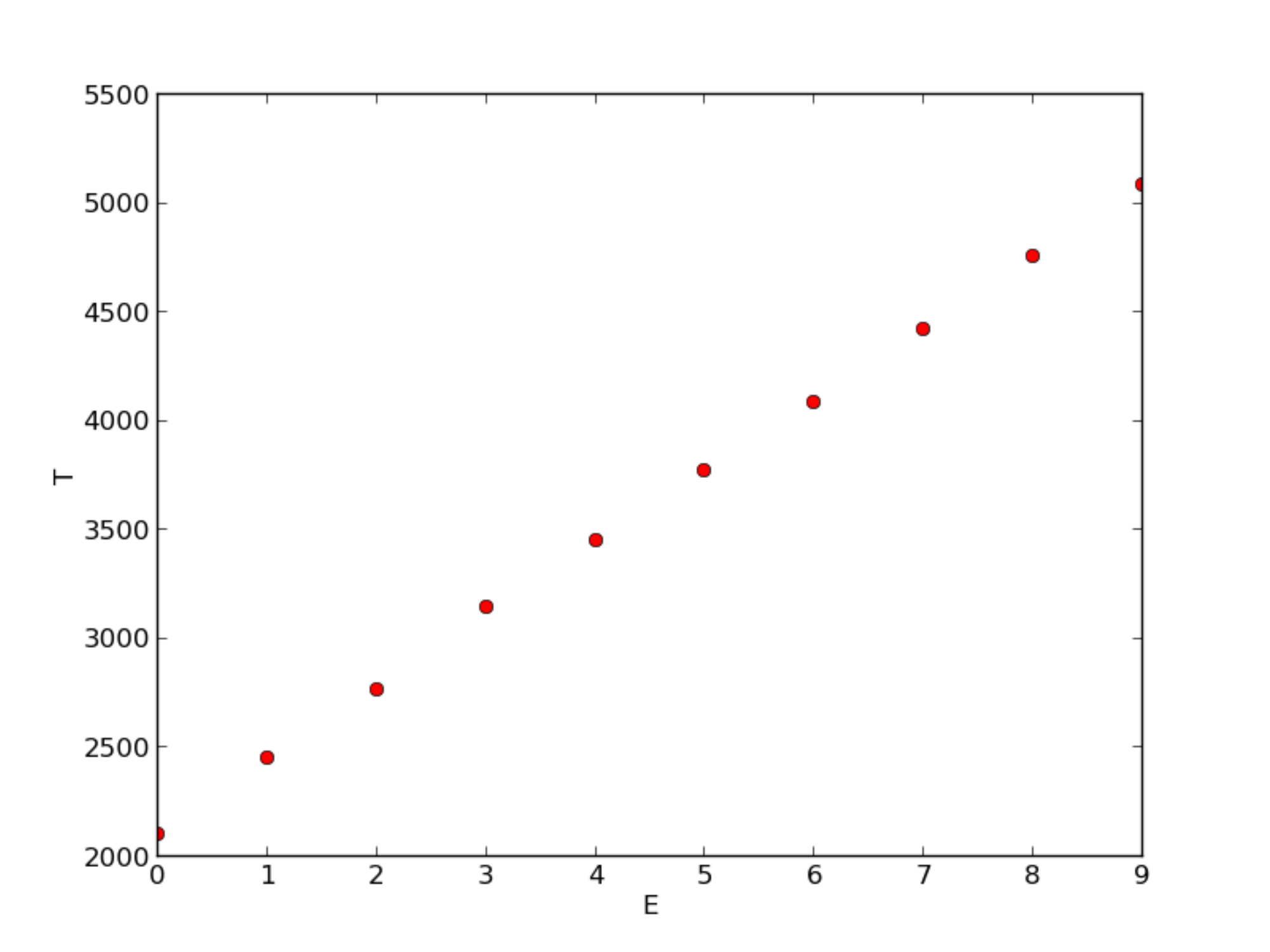} \\
\includegraphics[scale=0.4]{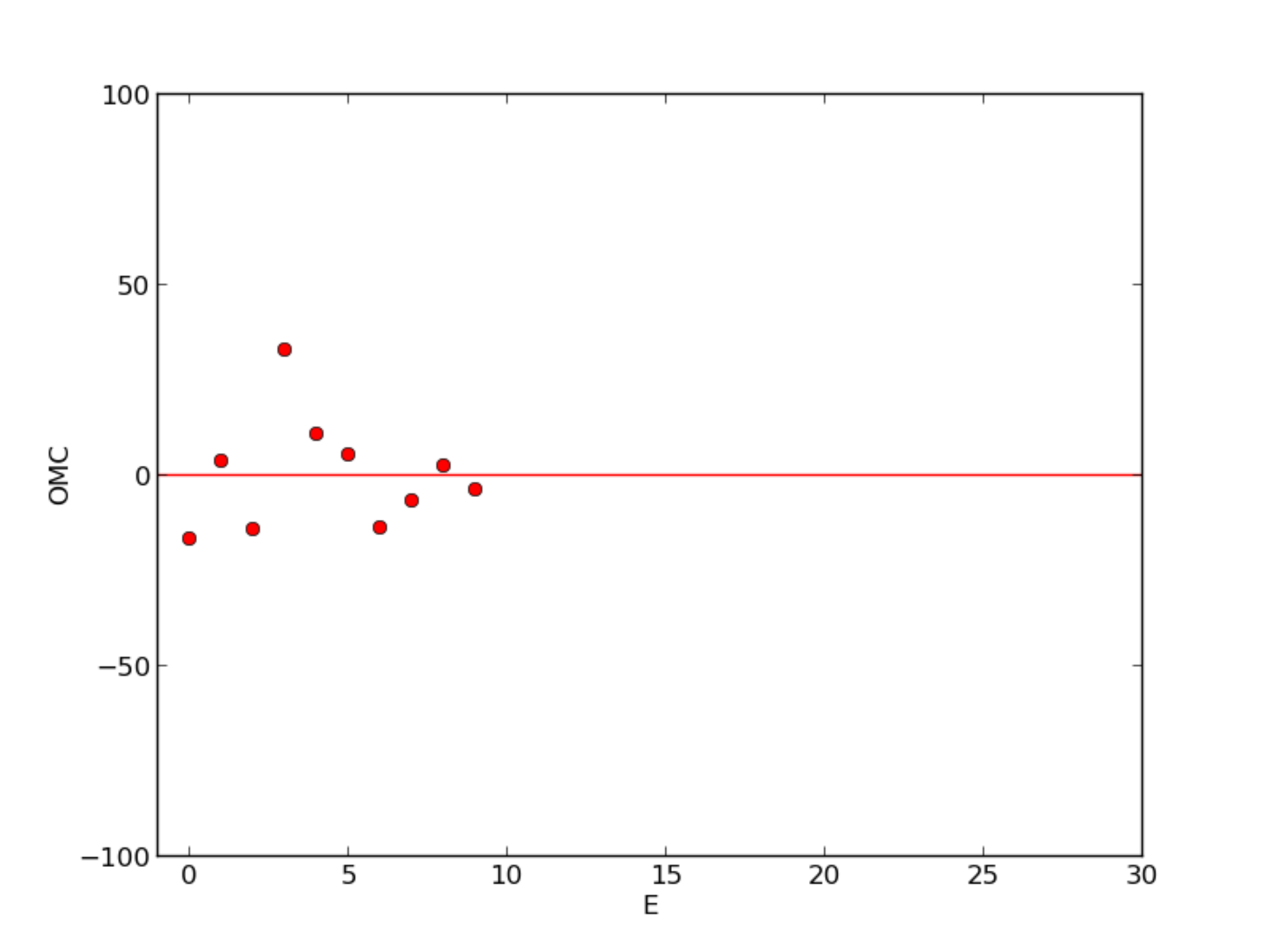} \\
\end{tabular}
\caption{The remaining diagrams of the example star Z Oct (see Figure 1), displayed in order to control the process: the peak magnitude E vs M$_{\rm max}$, (\textbf{upper panel}),  the moment of each maximum E vs T$_{\rm max,i}$ (\textbf{central panel}) and the standard deviation from the linear fit E vs O-C (\textbf{lower panel}).\label{Fig2}}
\end{figure}

In order to extract the crucial information from the ASAS database we developed a PYTHON code which determines the epochs and magnitudes of the maxima in their light curves, assigns correct cycle count numbers and derives the most important ephemeris data of each star. This procedure is not fully automatic; it requires interactive human collaboration. In this way, we could analyze 99.4\% of all stars listed in ASAS as Miras. 

Our method is illustrated with ASAS 174436-8629.2 (Z Oct) as a typical example. The first two steps of our procedure are shown in Figure~\ref{Fig1}. Firstly, the complete ASAS light curve is presented and the code selects a horizontal cut line which, by default, is placed at one third of the total amplitude (magnitude difference between the brightest and the faintest measurement in the data sample) below the brightest point of the data sample. Only values brighter than this limit are used in the period determination process. This line can be modified by the operator, in order to include fainter observations if some visible maxima are missed, or limit the analysis to brighter ones if isolated values are included which apparently do not present real light maxima.   

In the next step (Figure~\ref{Fig1}, lower panel) the code displays a histogram which shows the distribution of the time differences between subsequent data points in the light curve. In this example all observations with time differences $<$50 days belong to the same maximum while time differences around 270 days refer to jumps from one maximum to the next one. The PYTHON code identifies the first large gap in this distribution, and suggests a limit for the cycle counting process, if the two following conditions are fulfilled: the ratio between two adjacent time differences is $>$3.5 and the latter time difference is $>$10 days. These conditions have been determined via experimental checks in real data of several target stars. The cut (perpendicular red line in Figure~\ref{Fig1}) is then taken as the average between the largest epoch difference value of the first group and the smallest value of the second group. If this condition is not valid in any pair of time differences for a particular variable, the operator can enter the limit manually. The default limit can also be modified by the operator, if necessary. In this way, we obtain in most cases the correct assignment of cycle \textit{E} for all maxima, beginning with \textit{E} = 0 as the first one. The PYTHON code selects, within each group belonging to the same maximum, the brightest V magnitude value and assigns it as the magnitude at maximum light and the corresponding T$_{\rm max}$ as the moment of this maximum. A linear least-squares fit \textit{E} vs \textit{T$_{\rm max}$} leads to the period \textit{P}, the moment of the first maximum \textit{T$_{0}$}, the mean errors \textit{$\sigma$(P)}, \textit{$\sigma$(T$_{0}$)} and the standard deviation \textit{$\sigma$(T)} which corresponds to \textit{$\overline{O-C}$}. The code also calculates the average maximum brightness \textit{$\overline{M}_{\rm max}$} and its standard deviation \textit{$\sigma$($M_{\rm max}$)}. 

In Figure~\ref{Fig2} we show the remaining diagrams for our example case Z Oct. The operator checks these diagrams especially \textit{E} vs \textit{O-C}, which will reveal no-zero slopes and sudden discontinuities whenever the cycle counting procedure was not correct. In this case, the operator has the option to repeat the whole process varying some of the input parameters, until the result is satisfactory. If there are at least two well defined minima covered by the ASAS observations, the operator estimates also the magnitude at minimum light based on the light curve display (accuracy about 0.1 mag), and enters this information into the database. The amplitude \textit{A} is calculated subtracting the value of \textit{$\overline{M}_{\rm max}$} from the magnitude at minimum light. However, since for most stars their minimum phases are below the threshold of ASAS, amplitudes could be determined only for 855 of our targets.

\section{Data presentation}

This procedure leads to a total of 9 general parameters, listed for all stars in our first on-line Table and reproduced here in Table~\ref{Fig1} for arbitrary chosen examples. Each line begins with the ASAS identification and that of the General Catalogue of Variable Stars GCVS (\citealt{s07}), if available, and lists the parameters \textit{T$_{0}$}, \textit{$\sigma$(T$_{0}$)}, \textit{P}, \textit{$\sigma$(P)}, \textit{$\sigma$(T)}, \textit{$\overline{M}_{\rm max}$}, \textit{$\sigma$(M$_{\rm max}$)}, \textit{A} and \textit{N} which refers to the total number of light maxima observed.  All T values refer to HJD-2450000. Our second on-line Table (here  shown as examples in Table~\ref{table2}) gives \textit{E} vs. \textit{T$_{\rm max,i}$} and \textit{E} vs. \textit{M$_{\rm max,i}$}, the most important parameters of each individual light maximum, in order to enable observers to combine easily our data with those of other sources. In this way, we could analyze the data of a total of 2875 stars classified as Mira in the ASAS catalogue leaving only 20 stars that we were not able to analyse with our method. They are listed in Table~\ref{table3}, and in each case the reason of elimination is given. 

\begin{table*}
\centering
\caption{Parameters of 3 examples of Mira stars, as determined with our method (first on-line data set).}
\label{table1}
\begin{tabular}{lllcccccccr}
\hline \hline 
ASAS name  & Alternative name  & T$_{0}[d]$ & $\sigma$(T$_{0}$)[d] & P[d] & $\sigma$(P)[d] & $\sigma$(T)[d] & $\overline{M}_{\rm max}$[mag] & $\sigma$(M$_{\rm max}$)[mag] & A[mag] & N \\
\hline 
000006+2553.2 & Z~Peg  & 2657.4 &  15.1 &  316.35 &  3.65 & 24.1 & 8.63  &  0.67 & 3.67 &  7 \\
000017+2636.4 & AH~Peg & 2910.7 &  8.1  &  374.23 &  2.19 & 11.6 & 11.21 & 0.36 & \nodata  & 6 \\                 
000837-3913.2 & V~Scl  & 1934.4 &  4.7  &  297.41 &  0.88 & 7.5  &  9.17 & 0.20 & 5.03  & 7\\
\hline
\end{tabular}
\end{table*}

\begin{table*}[t]
\centering
\caption{Cycle count numbers E vs. T$_{\rm max,i}$ (HJD - 2450000) and E vs. M$_{\rm max,i}$ (V magnitude) at each individual maximum of all stars in on-line Table 1 (second on-line data set). Examples referring to the 3 stars in Table \ref{table1}.}
\label{table2}
\begin{tabular}{lllllllllllllllll}
\hline \hline 
ASAS Name & Other Name   & E & 0 & 1 & 2 & 3 & 4 & 5 & 6 & 7 & 8 & 9 \\
\hline 
000006+2553.2 & Z~Peg  &	T$_{\rm max,i}$ & 2629.5    &	2983.5    &	3295.6   &	3617.7    &	3912.9    &	4273.9    & \nodata & \nodata  &	5164.6 & \nodata \\ 		
&	      &	M$_{\rm max,i}$ & 9.51 	 &	8.32 	   &	8.04      &	7.82       &	9.30 	   &	8.31 	   & \nodata & \nodata &	 9.10 & \nodata \\
000017+2636.4 & AH~Peg  & T$_{\rm max,i}$ & 2917.6    &	3270.6    &	3662.6    &		\nodata    & 4409.5      &	4794.5    &	5145.6 & \nodata & \nodata & \nodata \\
&	      &	M$_{\rm max,i}$&11.40 	 &	10.73 	   &	10.95     &	     \nodata      & 11.61        &	11.54      &	11.00 & \nodata & \nodata & \nodata \\
000837-3913.2&    V~Scl  & T$_{\rm max,i}$ & 1928.5    &	2235.5    &	2529.7    &	  \nodata      & 3133.9      &	       \nodata & 3707.6    &\nodata & 4316.7       &	4610.9    \\
&	      & M$_{\rm max,i}$&	9.22    &	8.93       &	9.36      &	     \nodata      &  9.22        & \nodata &	8.86   & \nodata & 9.26          &	9.34  
 \\
\hline
\end{tabular}
\end{table*}

\section{Statistical analysis}

\subsection{Comparison to Richards et al.(2012) and the VSX}

Our results of pulsation periods, V magnitudes and amplitudes can now be compared with those determined by \citet{r12}; in this work the authors have obtained a variability type classification of all 50122 ASAS variables in a fully automated way, using a machine learning procedure, and deriving also some of the above mentioned parameters. In addition, we compared our results with the parameters given the AAVSO International Variable Star Index VSX which is the most comprehensive catalogue of variable stars the Milky Way available, with actually more than 398000 entries.

\begin{figure}[hbtp]
\centering
\begin{tabular}{cc}
\includegraphics[scale=0.45]{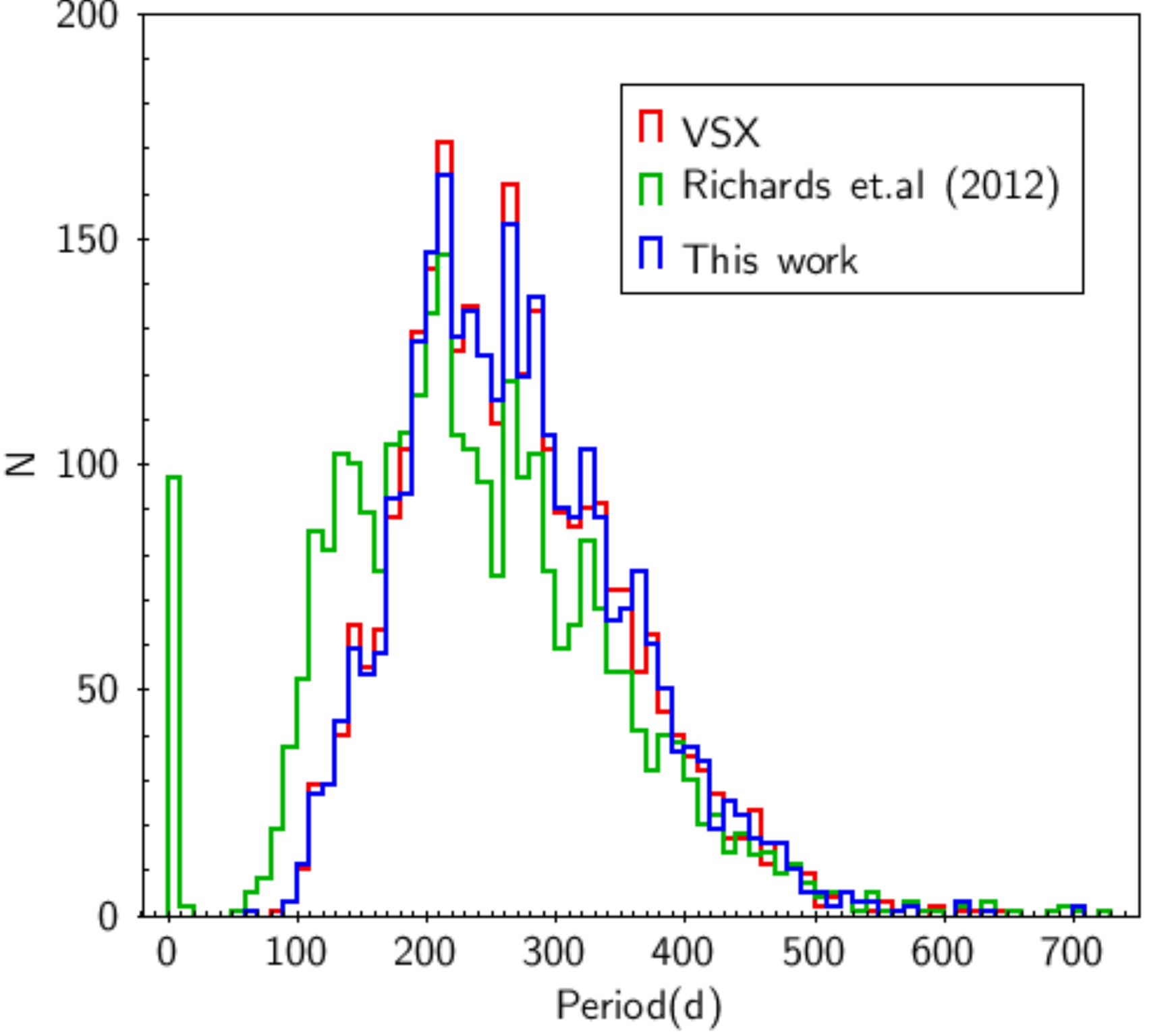} \\
\includegraphics[scale=0.45]{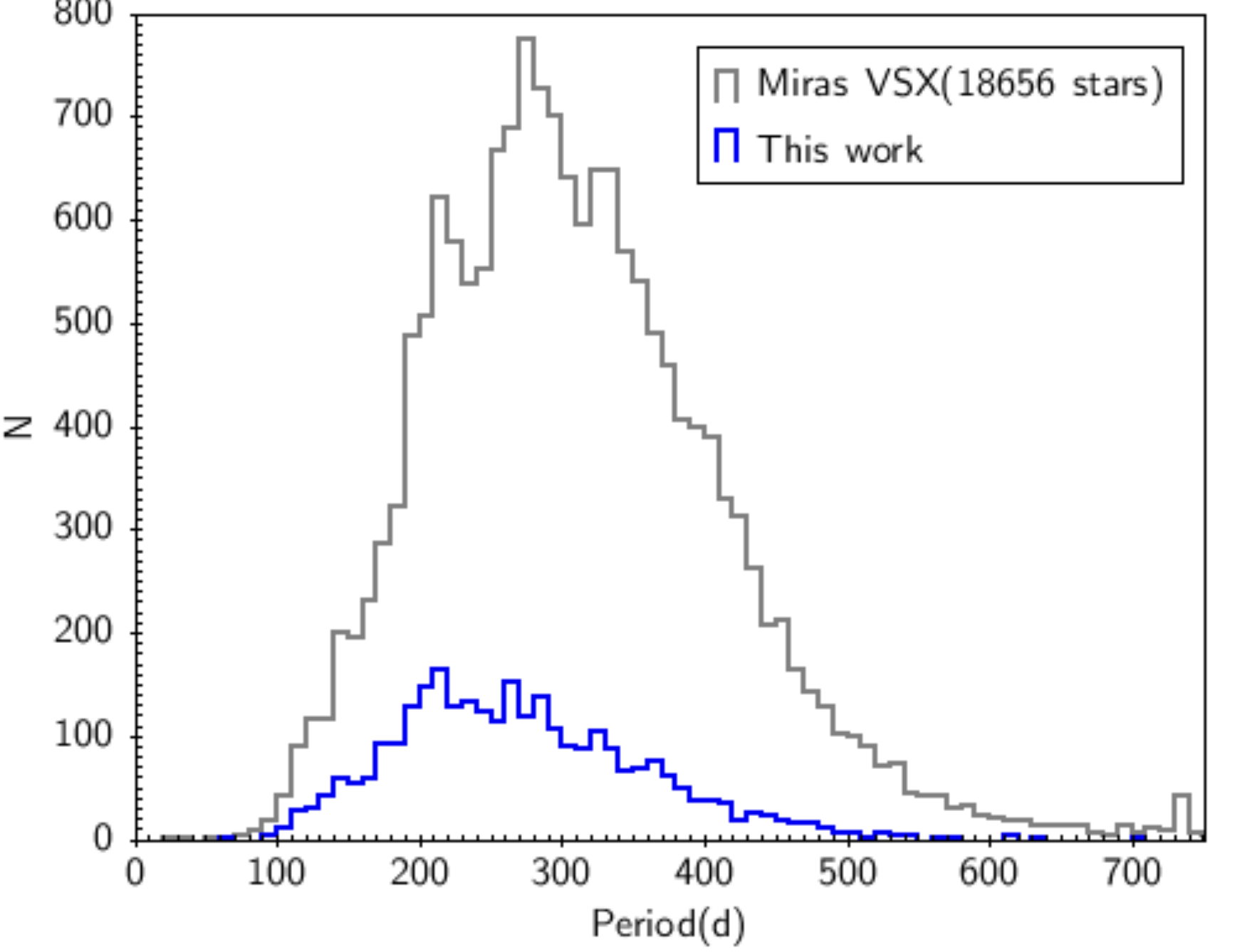} \\
\end{tabular}
\caption{Comparisons of different period of distributions. The binning of all histograms is 10 days. \textbf{Upper panel}: Period distribution of our results (blue), compared to those of Richards et al (2012, green) and of the VSX stars common with our Table 1 (red). \textbf{Lower panel}: Period distribution of all Mira stars with known period listed in the VSX (grey), compared to our distribution (blue).\label{Fig3}}
\end{figure}

The period distributions of stars in common belonging to these two databases are shown in Figure~\ref{Fig3} (upper panel). The mean value of our periods is $\overline{P}$ = 271.1d $\pm$ 1.4 (cf. Table~\ref{table6}), while for the VSX $\overline{P}$ = 269.5d $\pm$ 1.6, based on 2868 common stars (excluding an extreme period of 5000 days in VSX). Therefore, these both mean periods agree within the expected margin. On the other hand, we found a considerably shorter mean period in Richards et al (2012) $\overline{P}$ = 241.1d. This includes 99 stars with  $\sim$1-day period values which apparently refer to cases whose periods remained undetermined by the machine learning procedure of \citet{r12}, resulting in a value that corresponds to the day-night rhythm at Las Campanas Observatory. Excluding these cases, we derived  $\overline{P}$ = 249.2 days $\pm$ 2.1 for the remaining 2776 stars of \citet{r12}, which is still significantly lower than our value and that of the VSX mean periods.

In Figure~\ref{Fig3} (lower panel) we compare our distribution with that of all 18656 Mira stars with known period listed in the VSX. Three local maxima at periods around 215, 275 and 330 days are present in all histograms shown in Figure~\ref{Fig3}. A very similar period distribution was already determined by \citet{v80} whose figure 8 shows a  histogram based on about 5000 Mira stars with known periods listed in the GCVS at that epoch. To our knowledge, there is no reference in the literature on this three-peak distribution of Mira periods.  

\begin{figure*}[hbtp]
\centering
\begin{tabular}{cc}
\includegraphics[scale=0.36]{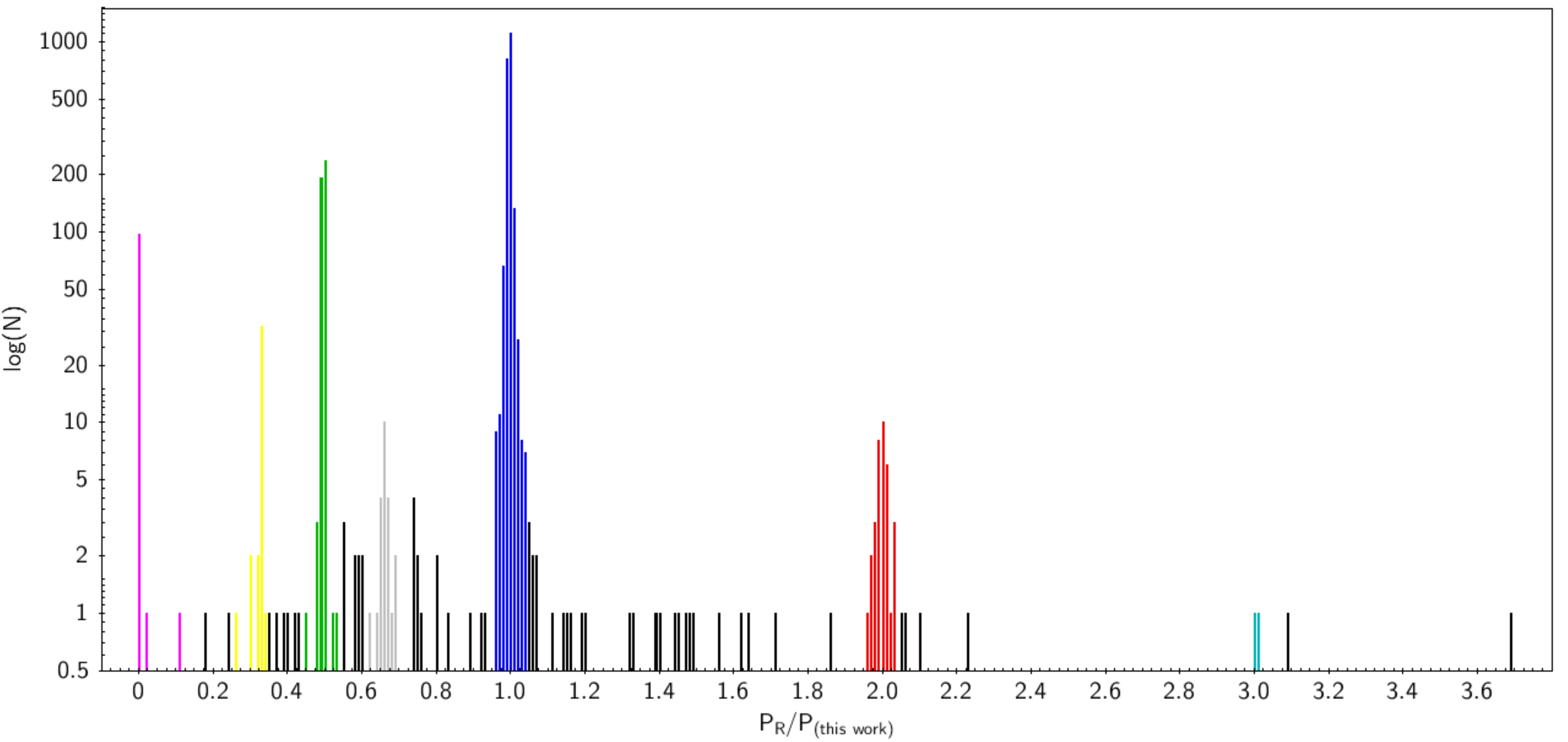} \\
\includegraphics[scale=0.36]{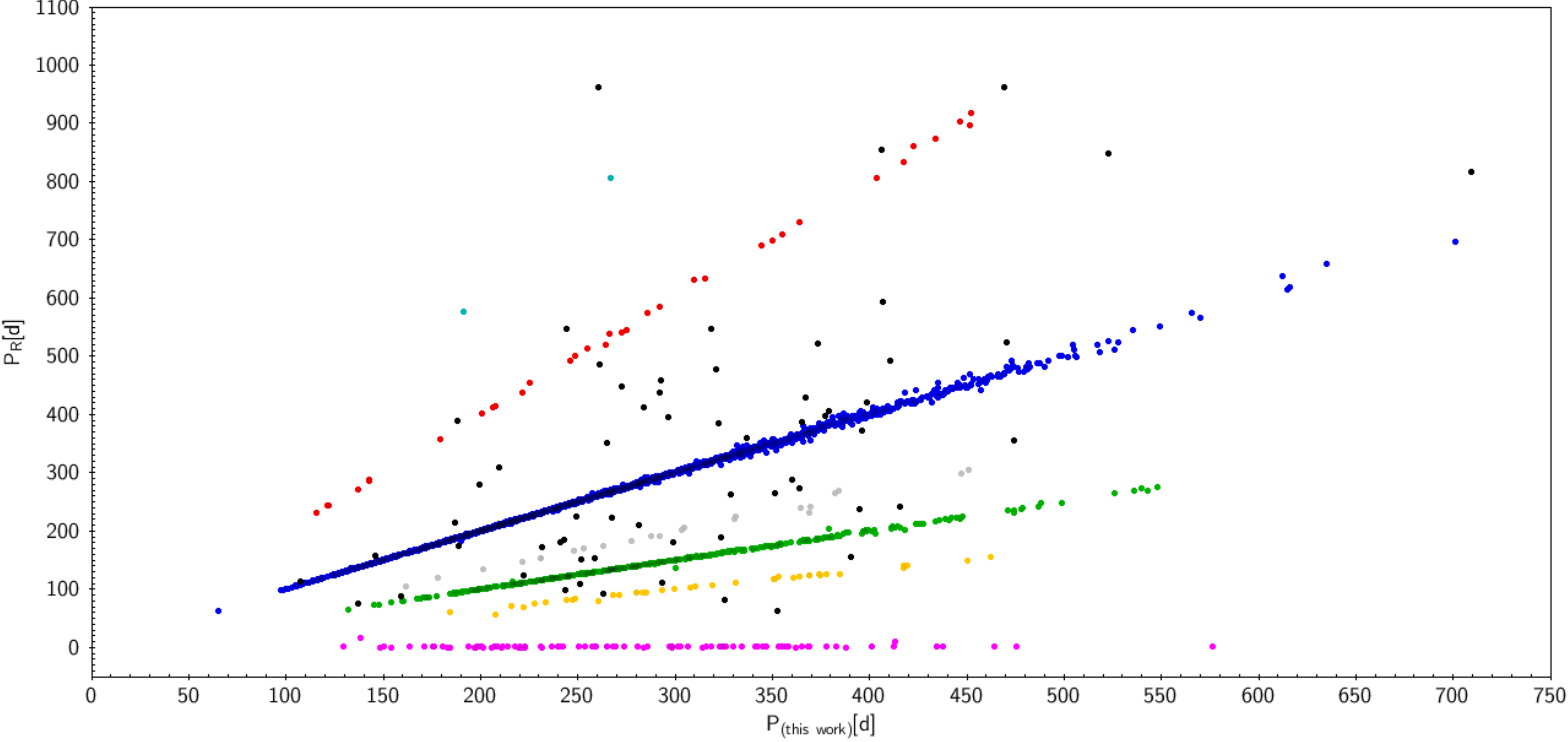} \\
\end{tabular}
\caption{Comparison between our periods and those by Richards et al. (P$_{R}$). The colors represent the different ratios P$_{R}$/P = 0:1 (pink), 1:3 (yellow), 1:2 (green), 2:3 (grey), 1:1 (blue), 2:1 (red) and 3:1 (light blue). \textbf{Upper  panel}: Histogram of the period ratios P$_{R}$/P. The black dots refer to cases with period ratios which do not coincide with the above mentioned values. \textbf{Lower panel}: Correlation between P and P$_{R}$.\label{Fig4}}
\end{figure*}

Figure~\ref{Fig4} (upper panel) shows a histogram of period ratios between Richards et al and our results. Several whole-number period ratios can be identified and are marked with different colors; the corresponding numbers are given in Table~\ref{table4}. Considering the correlation between our periods and those of Richards et al. (Figure~\ref{Fig4}, lower panel) we found significant differences to our results. Most of the  whole-number ratios should be aliases of the true period.  Only about 76\% of Richard's et al's periods are in agreement with our values. 

\begin{deluxetable}{llc}
\tabletypesize{\scriptsize}
\renewcommand\arraystretch{}
\tablecaption{Eliminated stars classified as Mira in ASAS (not listed in on-line Tables 1 and 2).\label{table3}}
\tablewidth{0pt}
\tablehead{
\colhead{ASAS name} & \colhead{Other name} & \colhead{Remarks:} \\
\colhead{}		    & \colhead{}			  & \colhead{(reason for elimination)}
}
\startdata
020751+2756.2 	& Z Tri 			& (1)     \\
035003-5721.2 	& RY Ret 	 	& (2)     \\      
042757+1602.6 	& W Tau  		& (2)     \\ 
054731+2708.3 	& AW Tau 		& (1)     \\
054957-5252.1 	& -      		& (2)     \\
060401+2113.5 	& V342 Ori 		& (2)     \\
070722+2818.0 	& AM Gem 		& (1)     \\ 
145641+2730.4 	& NSV 06861 		& (1)     \\ 
153557-4930.5 	& R Nor 			& (2)     \\
155042+1508.0 	& R Ser 			& (4)     \\
162229-4835.7 	& IO Nor 		& (3)     \\
172019-2309.7 	& -   			& (1)     \\
173808-2217.1 	& -   			& (1)     \\ 
174224-4344.9 	& RU Sco 		& (4)     \\
180444-2959.4 	& -     			& (1)     \\
183536-2409.2 	& -     			& (1)     \\ 
184041-1401.2 	& - 				& (1)     \\
190948+2804.3 	& TY Lyr  		& (1)     \\
193352+2819.6 	& TY Cyg  		& (1)     \\
204222+2728.8 	& EN Vul 		& (1)      
\enddata
\tablenotetext{Remarks:}{\\
(1) Only one or two maxima covered by ASAS \\
(2) Irregular light curve due to multiperiodicity (see Table~\ref{table8}) \\
(3) Misclassified as Mira; RCB star according VSX    \\
(4) No maxima are covered by ASAS  due to seasonal gaps}
\end{deluxetable}

\begin{figure*}[hbtp]
\centering
\begin{tabular}{cc}
\includegraphics[scale=0.36]{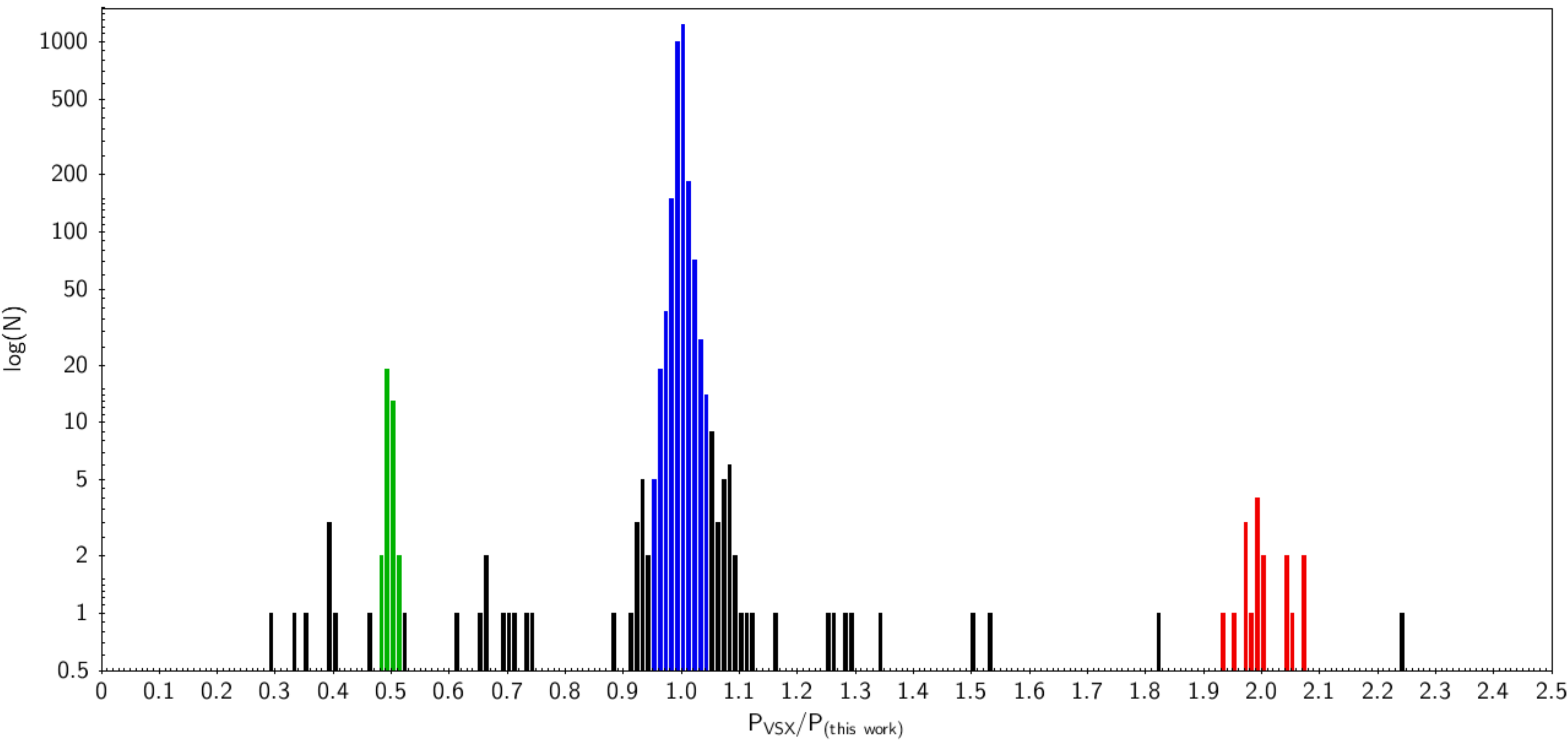} \\
\includegraphics[scale=0.36]{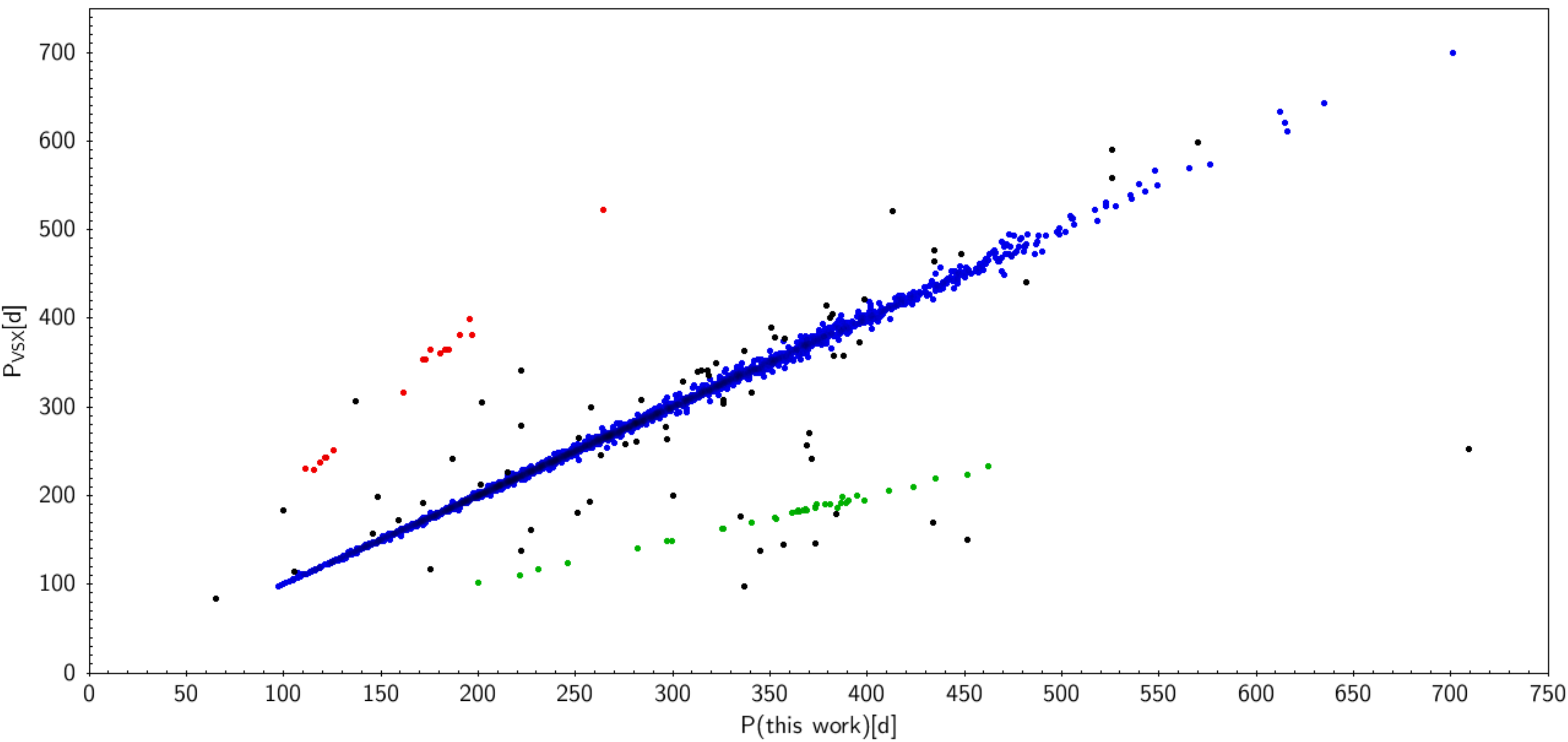} \\
\end{tabular}
\caption{Comparison of our periods with those of the VSX, in the same way as in Fig.4. \label{Fig5}}
\end{figure*}

This situation is rather different if we compare our periods with those of the VSX (Figure~\ref{Fig5}). There are also aliases, but at much lower extent, and only limited to half and double periods, while in more than 95\% of common stars our periods coincide with those of the VSX. The few aliases can probably be explained by an insufficient amount of observations (either in the VSX, in ASAS, or in both). The most important values of this comparison, as well as those with \citet{r12} are listed in Table~\ref{table4}.

\begin{figure}[hbtp]
\centering
\begin{tabular}{cc}
\includegraphics[scale=0.45]{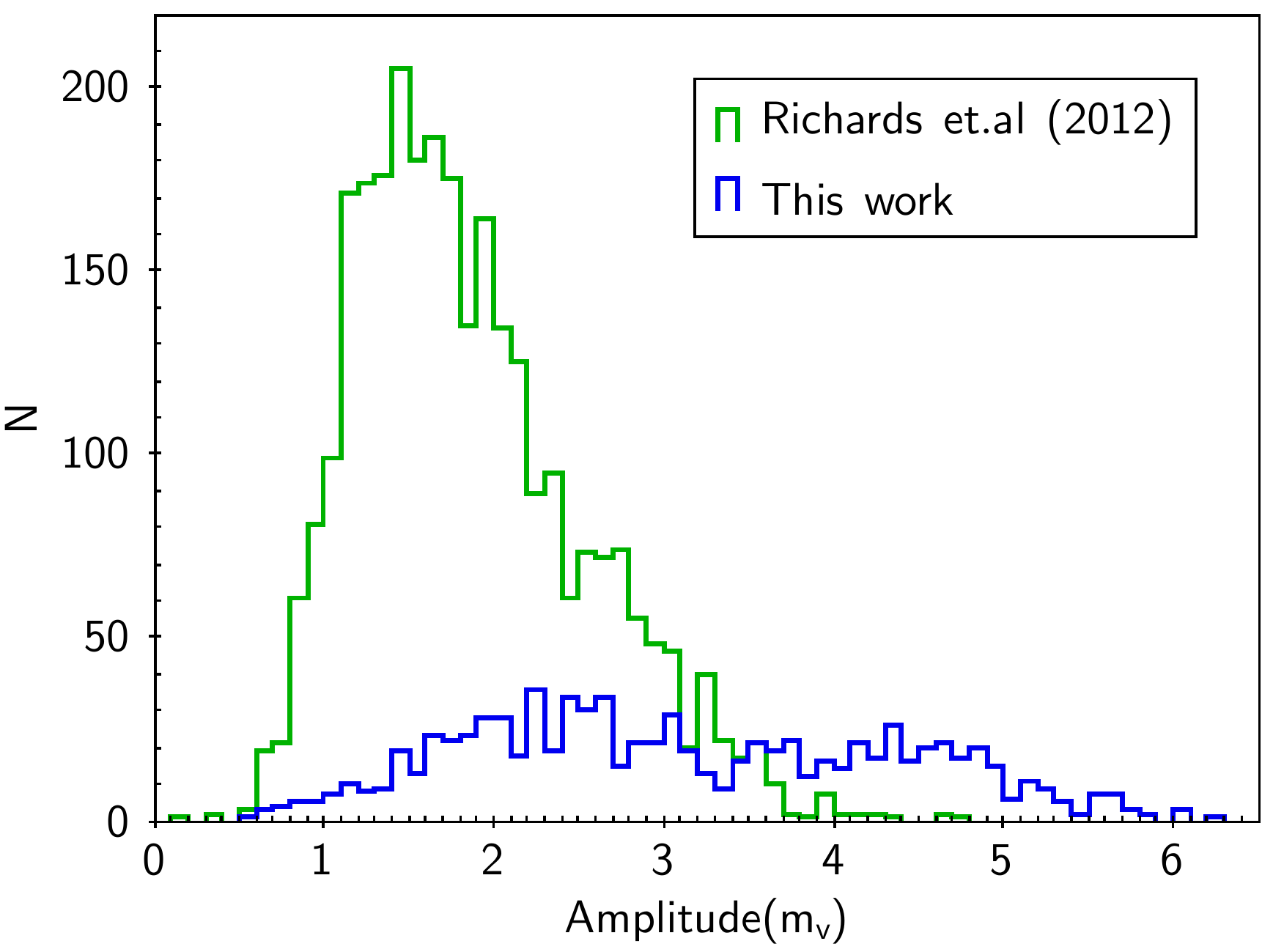} \\
\includegraphics[scale=0.45]{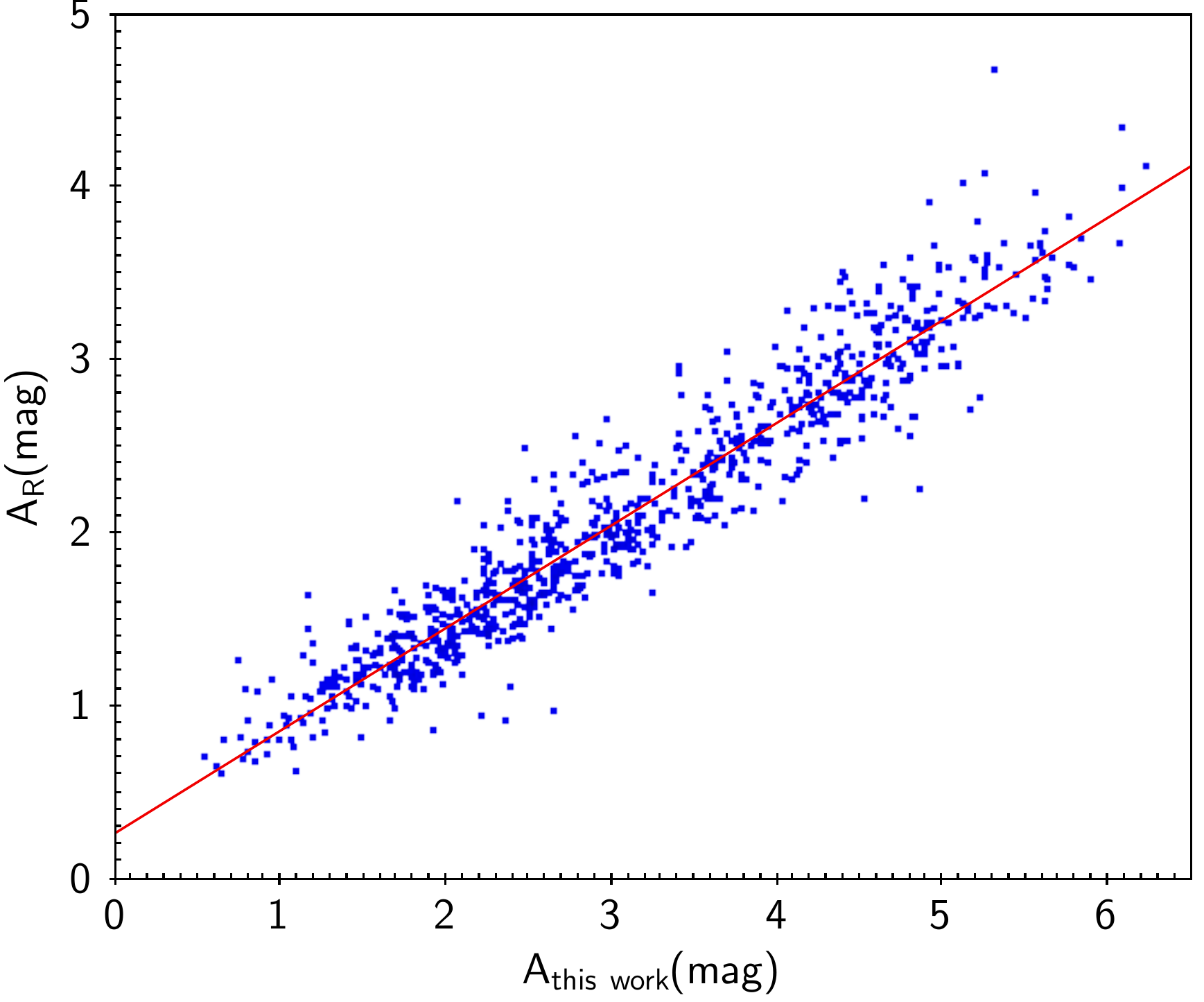} \\
\end{tabular}
\caption{\textbf{Upper panel:} Comparison of the amplitude distributions according to Richards et al (2012) and our results. Note the large excess of small amplitudes between 1 and 3 magnitudes in Richards et al (2012) which apparently is an artifact caused by their methods (see text). \textbf{Lower panel:} Correlation between our amplitudes and those of Richards et al (2012). The solid red line refers to a linear fit with the parameters given in Table 5.\label{Fig6}}
\end{figure}

In Figure~\ref{Fig6} (upper panel) we show the amplitude distributions. \citet{r12} determined amplitudes of all variable stars in ASAS, in an automated procedure, without considering that many targets could only be observed around their maxima by ASAS while the fainter phases of their light curves remained unobserved. This explains the large amount of small amplitudes with a peak at \textit{A} $\approx$ 1.4 mag, an artefact caused by the large number of faint targets in the sample. In Figure~\ref{Fig6} (lower panel) we restrict to the 855 stars in which we were able to determine the amplitude (i.e. whenever the ASAS light curve contains also the minimum magnitude level) and compare them with \citet{r12} as shown in Figure~\ref{Fig6} (lower panel). The slope, intercept and correlation coefficient of the corresponding least-squares fit is given in Table~\ref{table5}.

\begin{deluxetable}{cllllll}
\tabletypesize{\scriptsize}
\renewcommand\arraystretch{0.9}
\tablecaption{Period ratio distributions according to the comparison to Richards et al (2012) and to the VSX. “Other stars” refer to the black dots in Figures 4 and 5. N refers to star numbers.\label{table4}}
\tablewidth{0pt}
\tablehead{
\colhead{} & \multicolumn{3}{c}{Richards et al.(2012)}      & \multicolumn{3}{c}{VSX} \\
\colhead{} & \multicolumn{3}{c}{N$_{total}$=2875} & \multicolumn{3}{c}{N$_{total}$=2868} \\
\colhead{Period ratio} & \colhead{N} & \colhead{\%} & \colhead{limits} &
\colhead{N} & \colhead{\%} & \colhead{limits} }
\startdata
0:1 			& 99   & 3.4  & $<$0.15   & -    & -   &  -				\\ 	
1:3 			& 38   & 1.3  & 0.25-0.35 & -    & -   &  -				\\        
1:2 			& 435  & 15.1 & 0.45-0.55 & 36   & 1.3  &  0.475-0.525 \\
2:3 			& 23   & 0.8  & 0.62-0.72 & -    &	-  &	  -			\\
1:1 			& 2181 & 75.8 & 0.95-1.05 & 2752 & 95.9 & 0.95-1.05 	\\
2:1 			& 34   & 1.2  & 1.95-2.05 & 17   & 0.6  & 1.9-2.1     \\ 
3:1 			& 2	   & 0.1  & 2.95-3.05 & 	-    & -   & - \\
Other stars & 63   & 2.2  & -         & 63    & 2.2 & -   
\enddata
\end{deluxetable}

\begin{deluxetable}{llllc}
\tabletypesize{\scriptsize}
\renewcommand\arraystretch{}
\tablecaption{Linear fit parameters and correlation coefficients \textit{r}. (R) refers to Richards et. al. (2012). \label{table5}}
\tablewidth{0pt}
\tablehead{
\colhead{} & \colhead{Slope} & \colhead{Intercept} & \colhead{r} & \colhead{Reference} \\
\colhead{} & \colhead{} 		& \colhead{} 		  & \colhead{}  & \colhead{figure}
}
\startdata
A vs A(R) 							&	0.594 		&	0.266 		&		0.955	& 6 \\
M$_{\rm max}$ vs M$_{\rm max}$(R) 			&	0.950 		&	0.603 		&		0.963	& 8 \\
P vs $\sigma$(P) 					&	0.011 		&	-1.527 		&		0.657	& 9 \\
P vs $\sigma$(P)/P 					&	0.00001664 	&	0.000142 	&		0.451	& 9 \\
P vs A 								&	0.000860 	&	2.867 		&		0.0668	& 7 \\
P vs $\sigma$(T) 					&	0.02662 		&	5.254 		&		0.365	& 9 \\
P vs $\sigma$(T$_{0}$) 				&	0.0253 		&	0.968 		&		0.485	& 9 \\
P vs M$_{\rm max}$ 						&	-0.000252 	&	12.021 		&		-0.165	& 10 \\
P vs $\sigma$(M$_{\rm max}$)				&	0.000526 	&	0.199 		&		0.301	& 10 \\
M$_{\rm max}$ vs $\sigma$(M$_{\rm max}$) 	&	-0.0238 		&	0.611 		&		-0.208	& -
\enddata
\end{deluxetable}

\begin{figure}[hbtp]
\centering
\begin{tabular}{cc}
\includegraphics[scale=0.45]{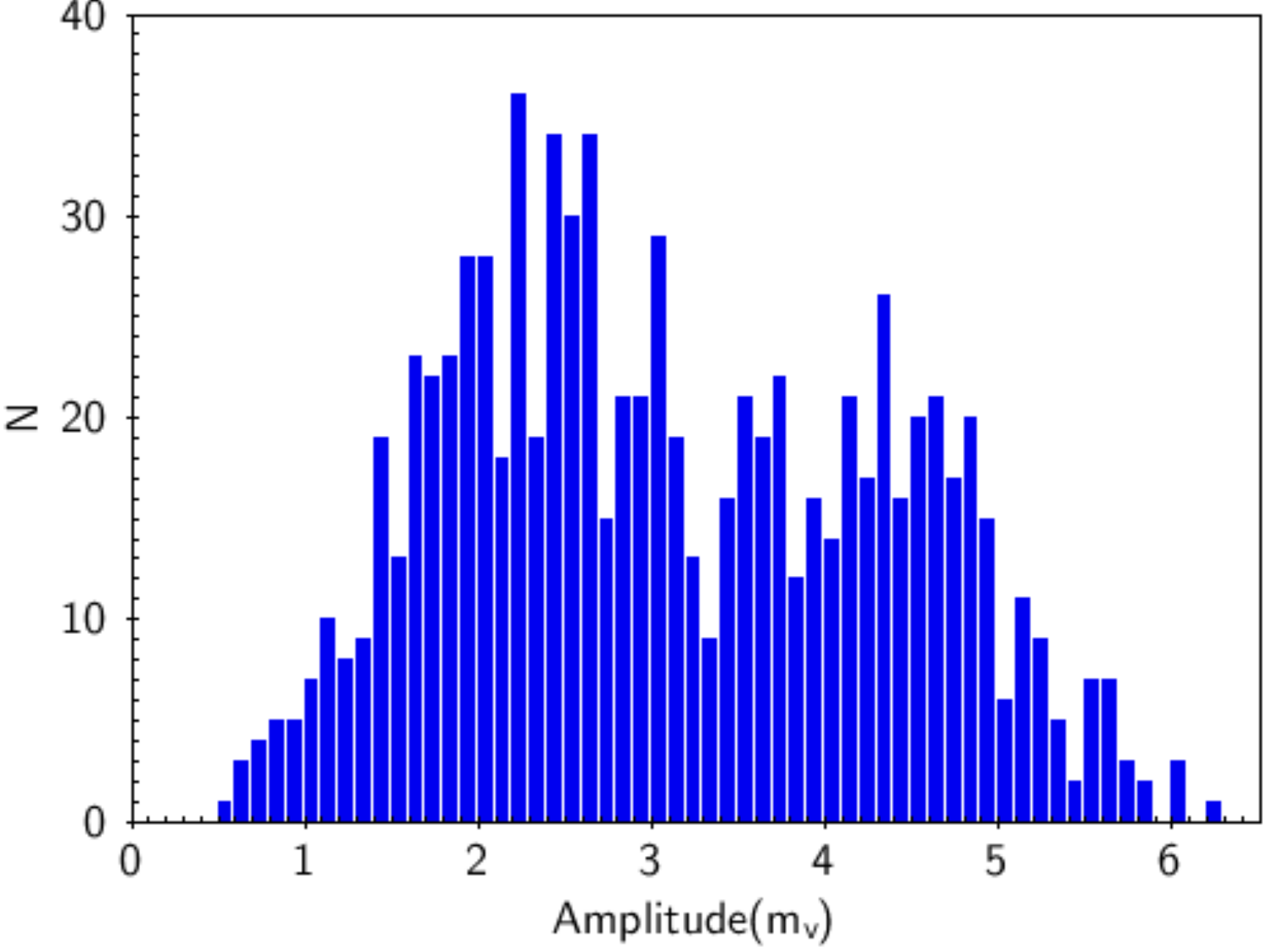} \\
\includegraphics[scale=0.45]{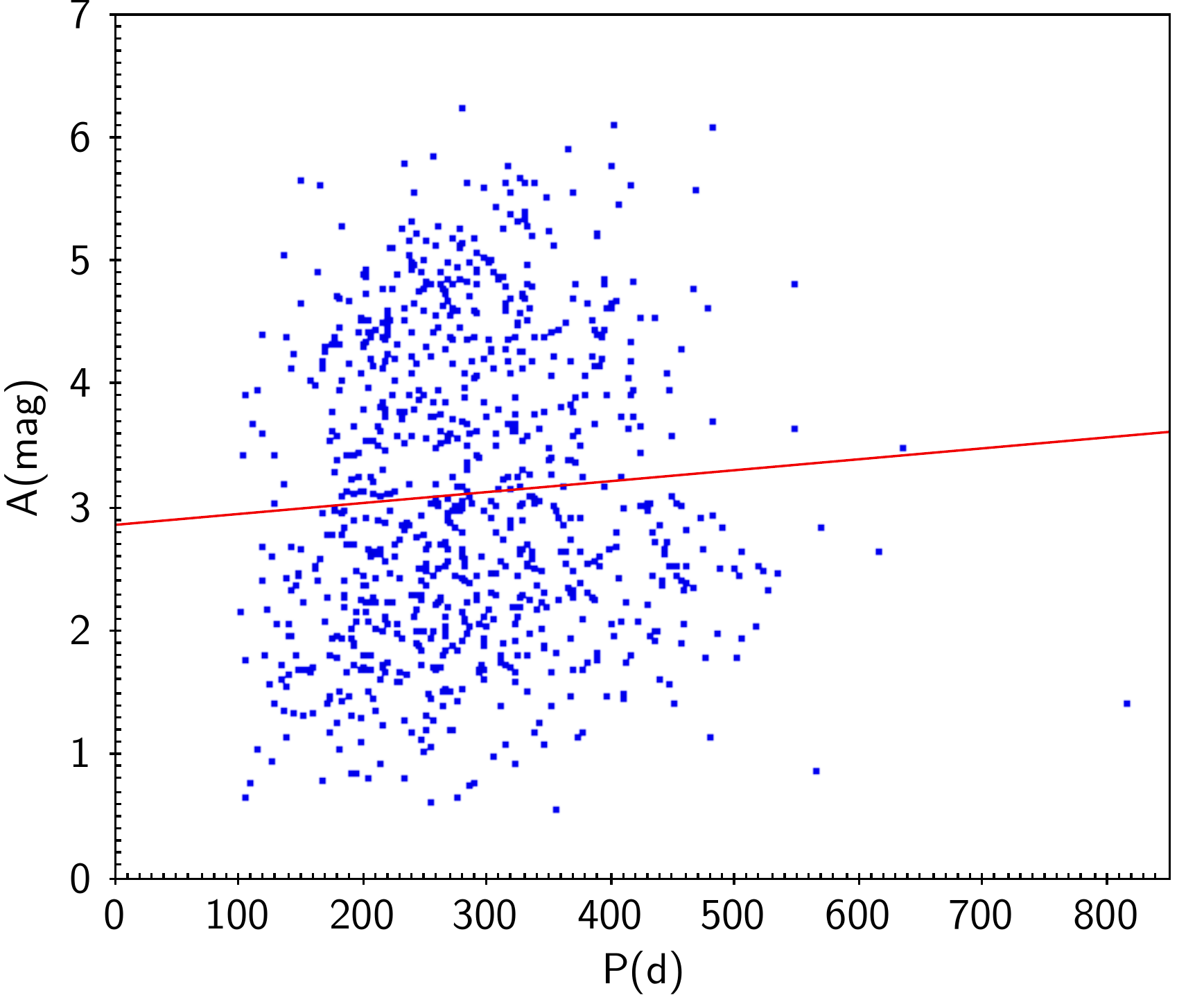} \\
\end{tabular}
\caption{\textbf{Upper panel:} Histogram of our amplitudes. There are indications for a bi-modal distribution, with maxima around 2.3 and 4.2$^{m}$ and a remarkable minimum near 3.3$^{m}$. \textbf{Lower panel:} Correlation between period and amplitude. The solid red line corresponds to the fit parameters given in Table 5.\label{Fig7}}
\end{figure}

Our amplitudes in Figure~\ref{Fig7} show a slight tendency towards a bimodal distribution with maxima at 2.5$^{m}$ and 4.5$^{m}$, and a deficit around 3.4$^{m}$. Apparently, the amplitudes are not correlated with the periods (Figure~\ref{Fig7}, lower panel). However, the amplitude distribution is severely affected by the presence of an SR population among the stars classified as Mira in ASAS. A total of 281 targets with known amplitude (33\%) have A $<$ 2.5$^{m}$, smaller than expected for Mira stars. From the remaining 2020 stars, 915 have \textit{$\overline{M}_{\rm max}$} $>$ 12.0$^{m}$, i.e. only 2.5$^{m}$ brighter than the ASAS limiting magnitude V = 14.5$^{m}$. Among them there could be some SR variables, while all other targets will have larger amplitudes, typical for Mira stars. If we adopt the same fraction of 33\% (as found in the sample with known amplitude) to be of SR type among the stars \textit{$\overline{M}_{\rm max}$}$>$12.0$^{m}$, we could expect 302 other SR variables in the 'faint sample'. This estimation leads to a total of about 583 SR stars (20\%) of the total sample investigated here. As an example we could mention BI Car which is the star with the largest period (817d) in our sample, with an amplitude of only A = 1.42$^{m}$; this star is classified as SRB in the GCVS.

\begin{figure*}[hbtp]
\centering
\begin{tabular}{cc}
\includegraphics[scale=0.45]{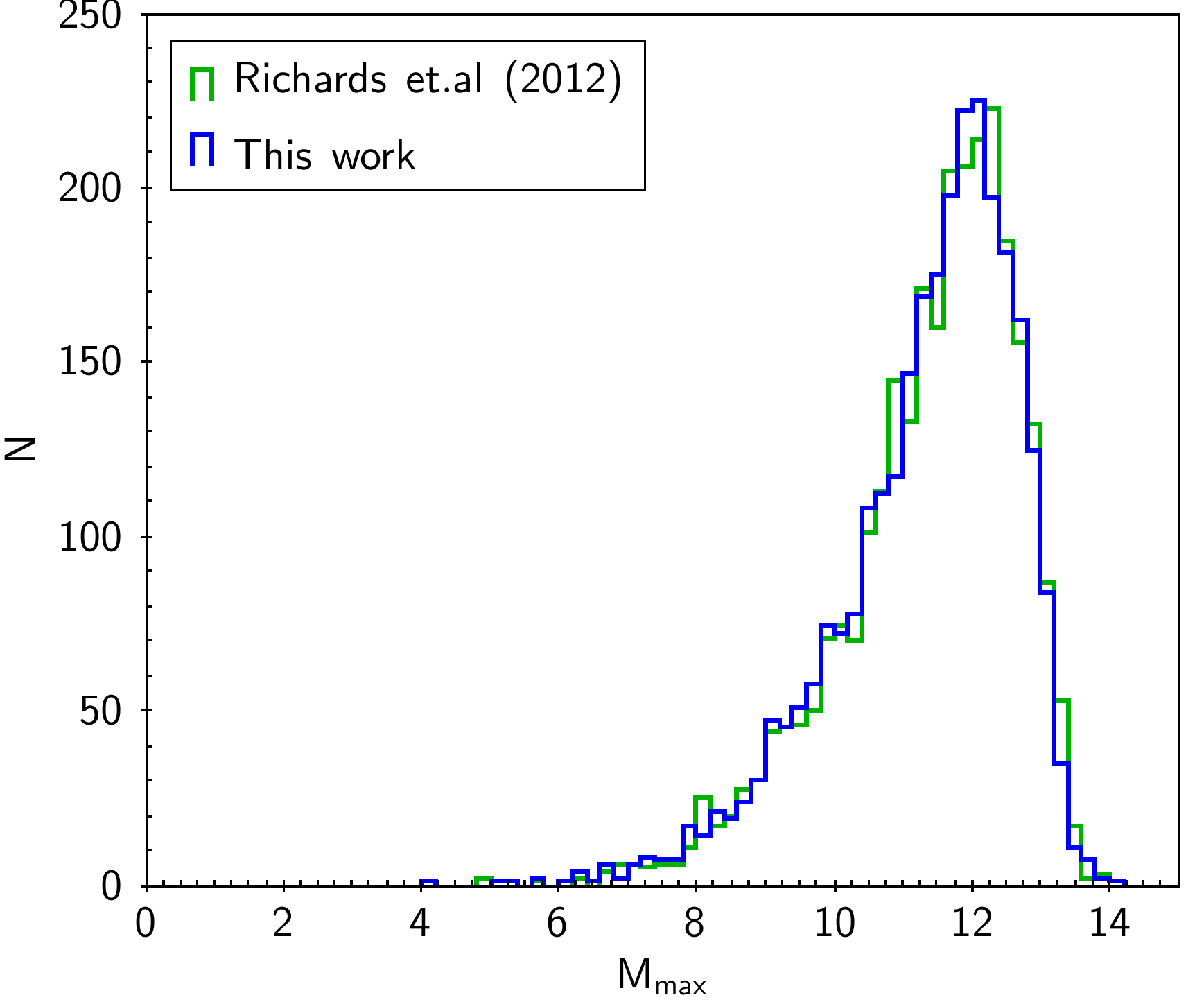} &
\includegraphics[scale=0.45]{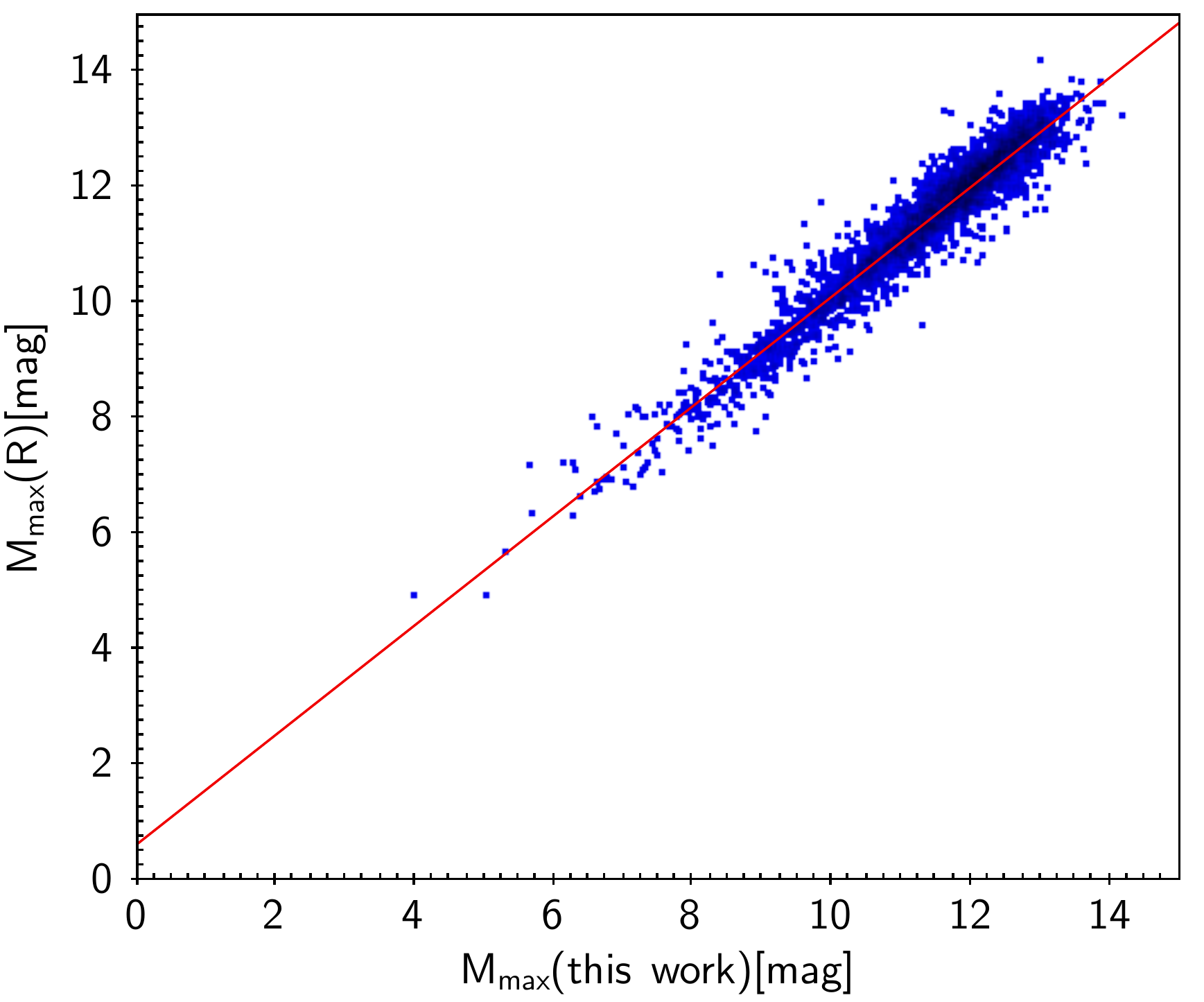} \\
\end{tabular}
\caption{\textbf{Left panel:} Comparison of the distributions of  the mean maximum brightness determined by Richards et al (2012) and according to our results (blue). \textbf{Right panel:} Correlation between our mean maximum brightness and that of  Richards et al (2012). The solid red line corresponds to the fit parameters given in Table 5.\label{Fig8}}
\end{figure*}

In Figure~\ref{Fig8} we compare the mean maximum magnitudes, were we find a perfect agreement with \citet{r12}. This is not surprising since \textit{$M_{\rm max}$} should not be affected by any of the known selection effects. 

\begin{figure*}[hbtp]
\centering
\begin{tabular}{cccc}
\includegraphics[scale=0.45]{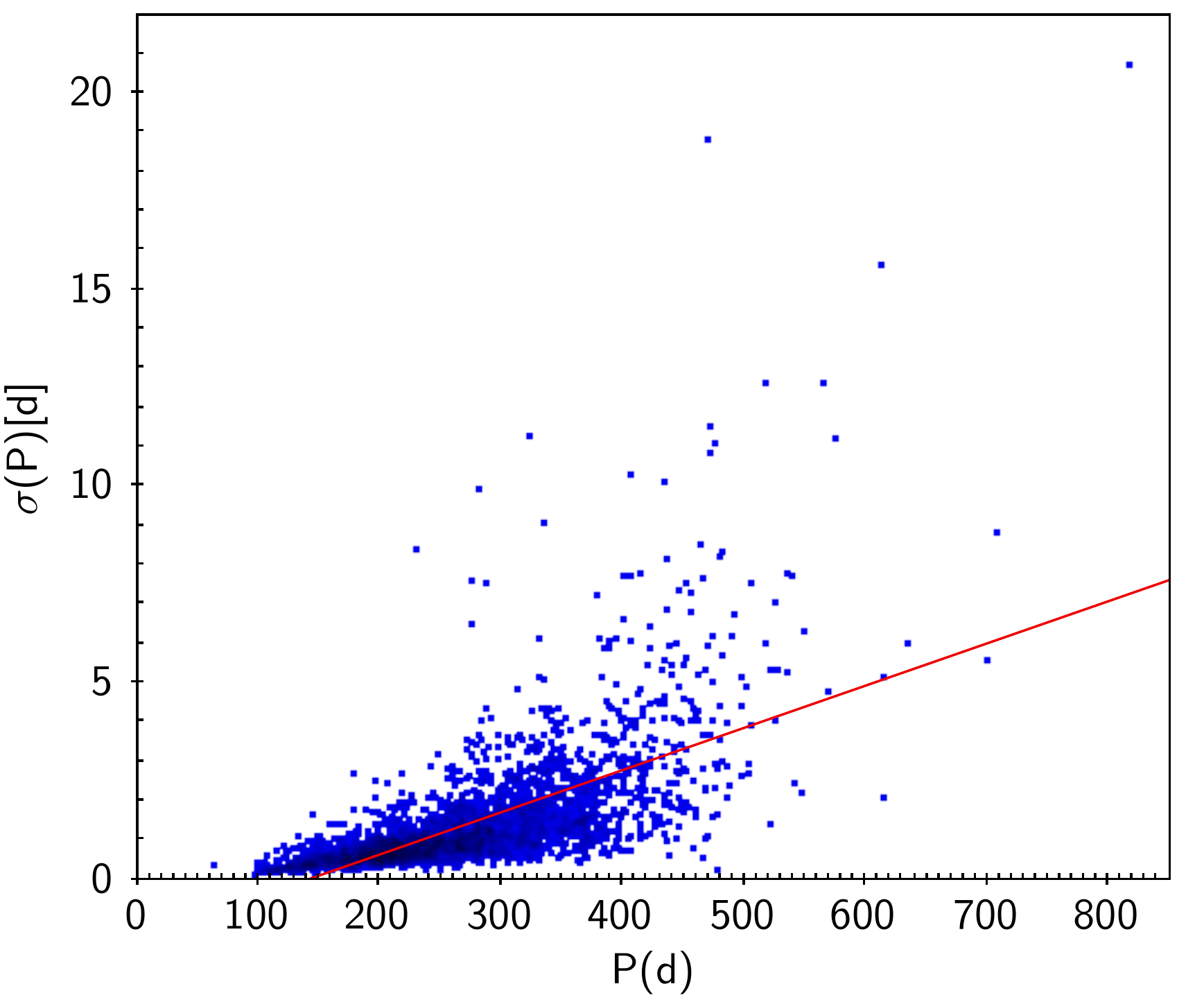} &
\includegraphics[scale=0.45]{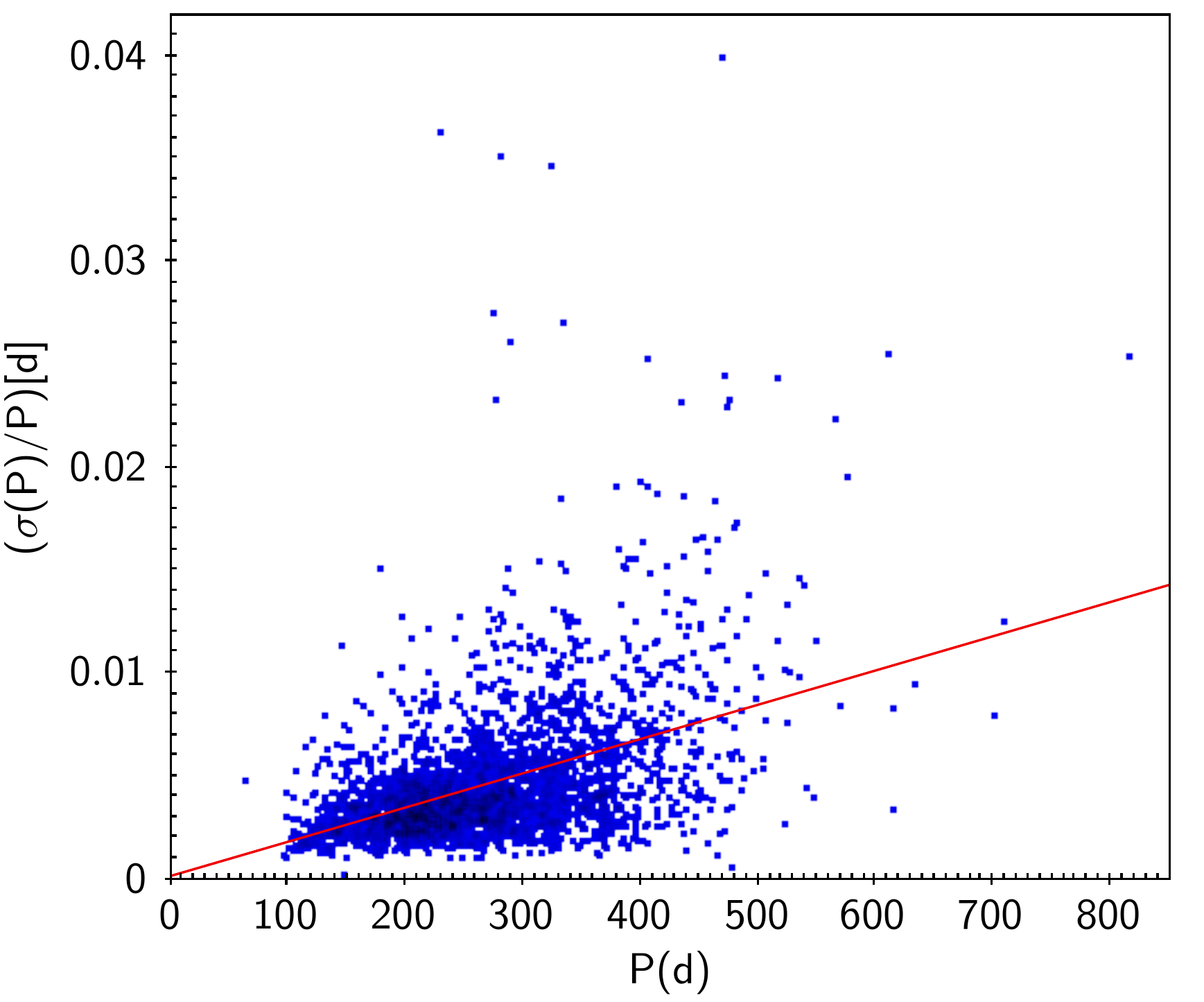} \\
\includegraphics[scale=0.45]{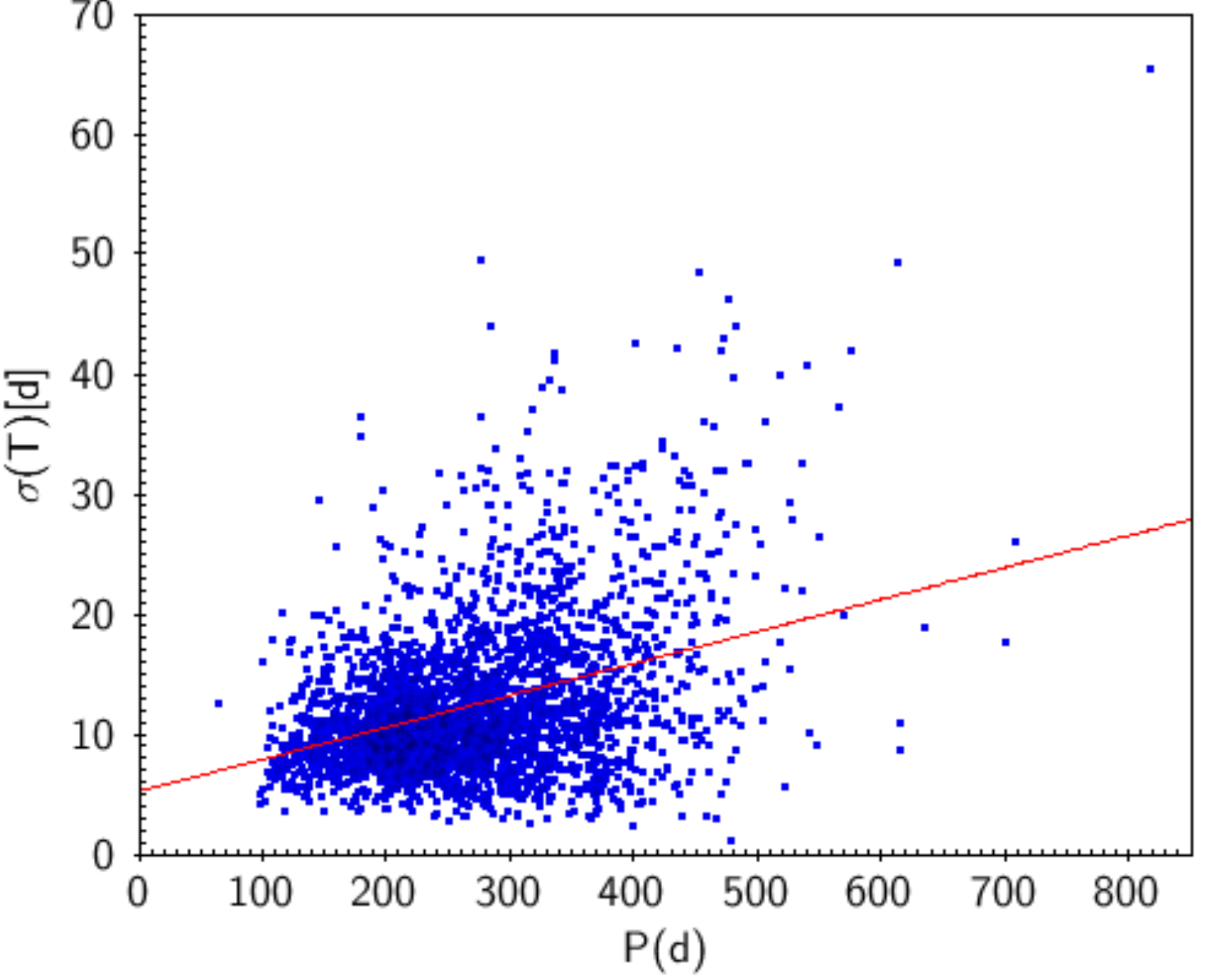} &
\includegraphics[scale=0.45]{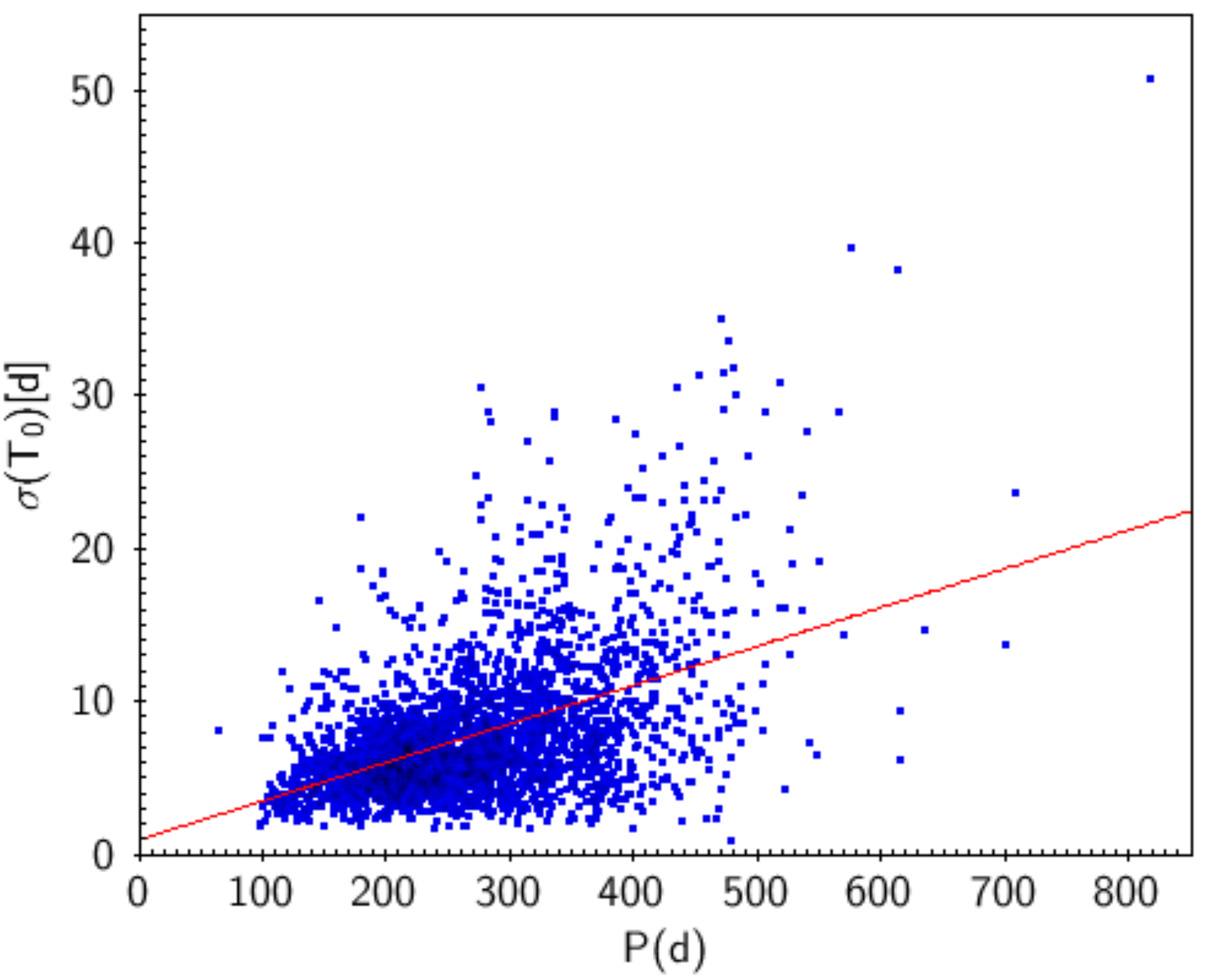} \\
\end{tabular}
\caption{Correlations between our period and the parameters $\sigma$(P), $\sigma$(P)/P, $\sigma$(T), and $\sigma$(T$_{0}$). The solid red lines correspond to the fit parameters given in Table 5.\label{Fig9}}
\end{figure*}

\begin{figure}[hbtp]
\centering
\begin{tabular}{ccc}
\includegraphics[scale=0.45]{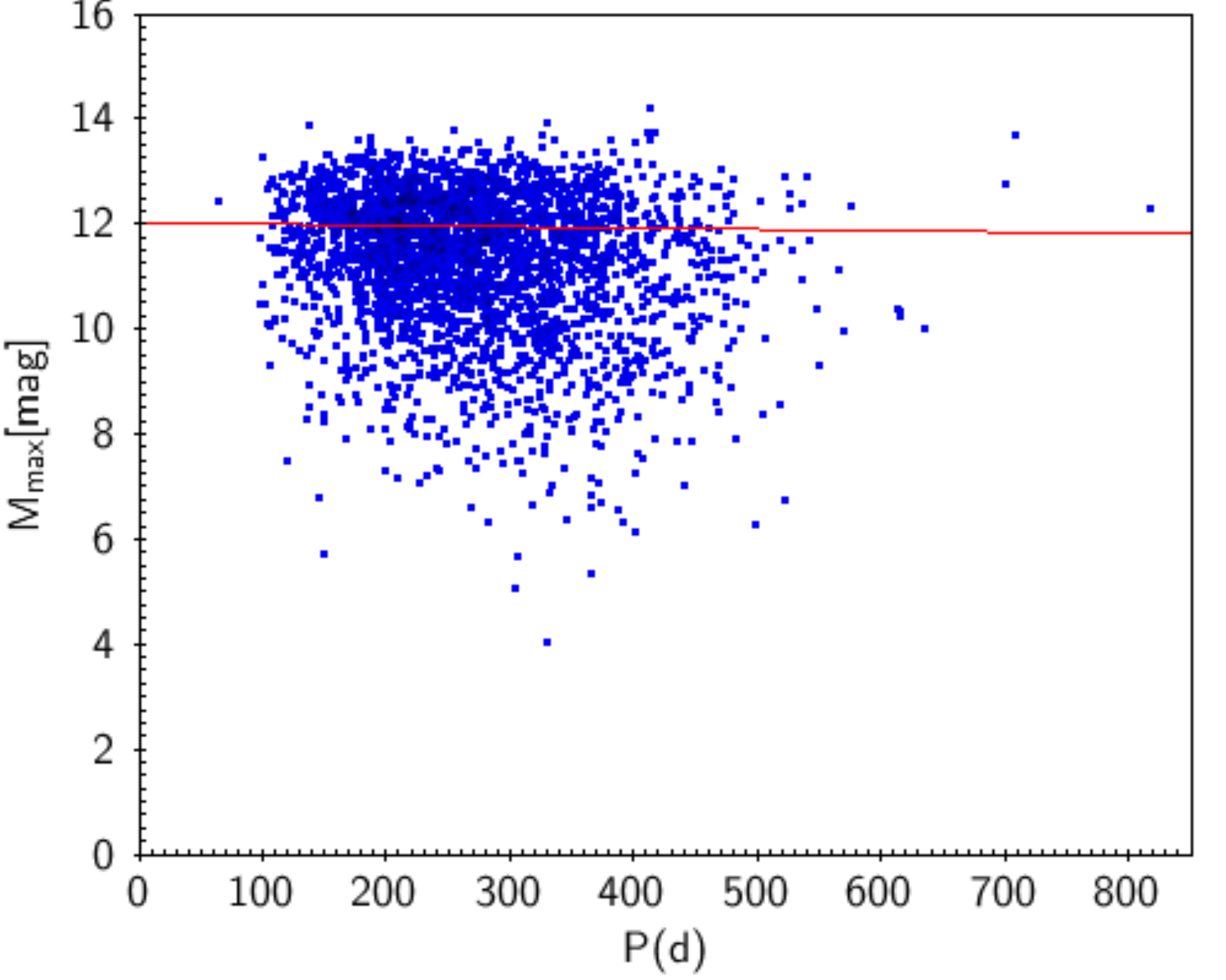} \\
\includegraphics[scale=0.45]{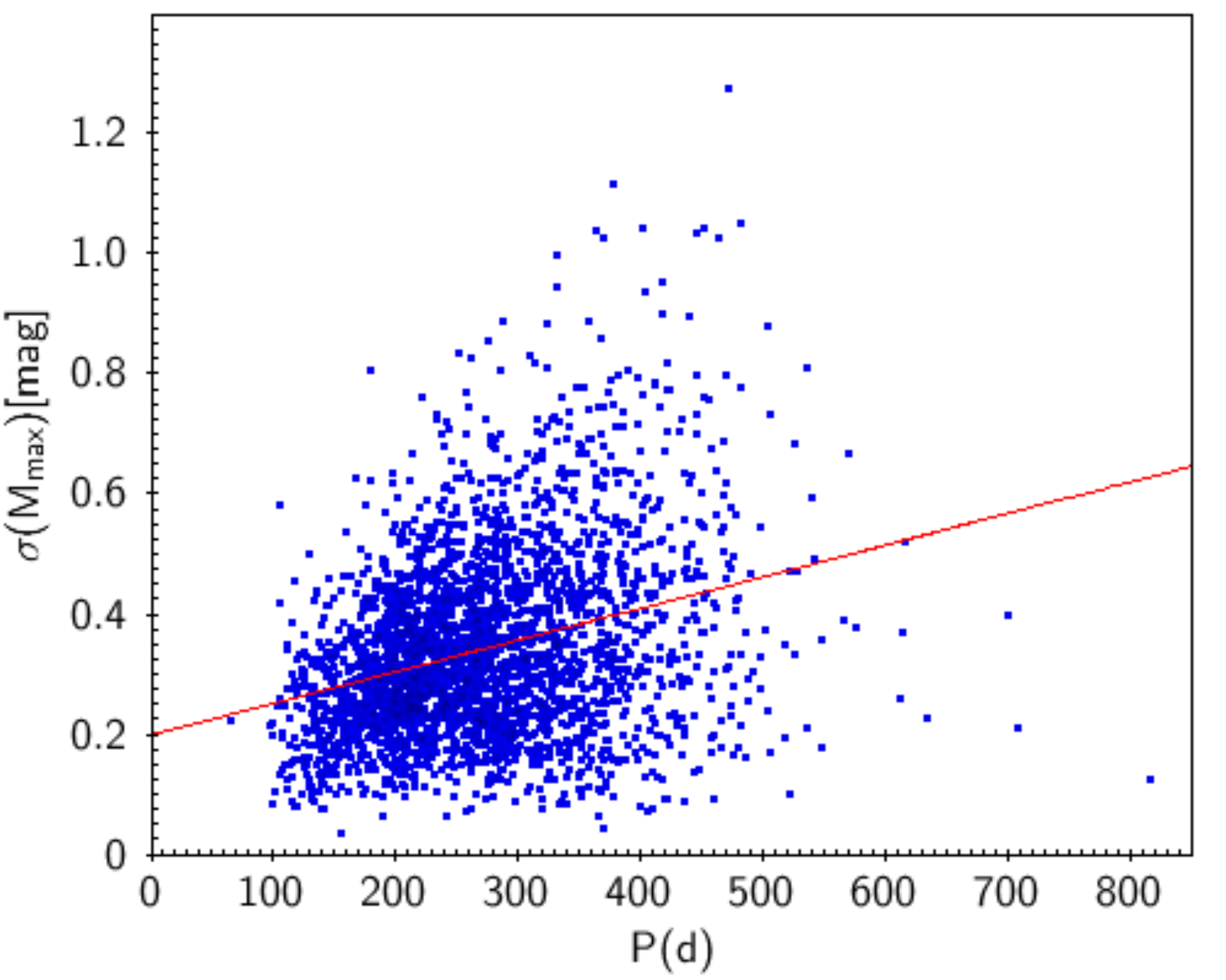} \\
\end{tabular}
\caption{Correlations between our period and the parameters M$_{\rm max}$ and $\sigma$(M$_{\rm max}$). The solid red lines correspond to the fit parameters given in Table 5.\label{Fig10}}
\end{figure}

\begin{figure}[hbtp]
\centering
\begin{tabular}{ccccc}
\includegraphics[scale=0.32]{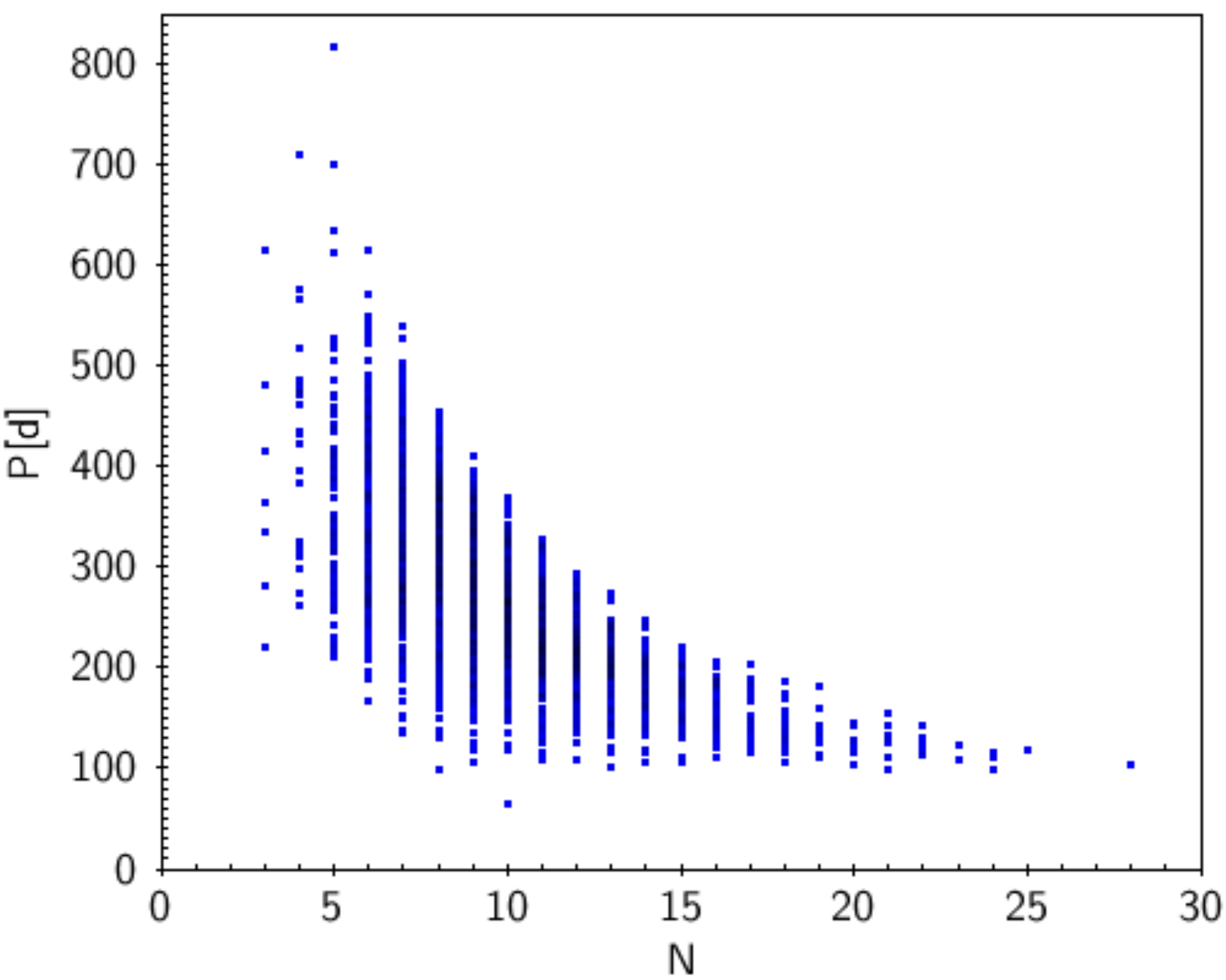} \\
\includegraphics[scale=0.32]{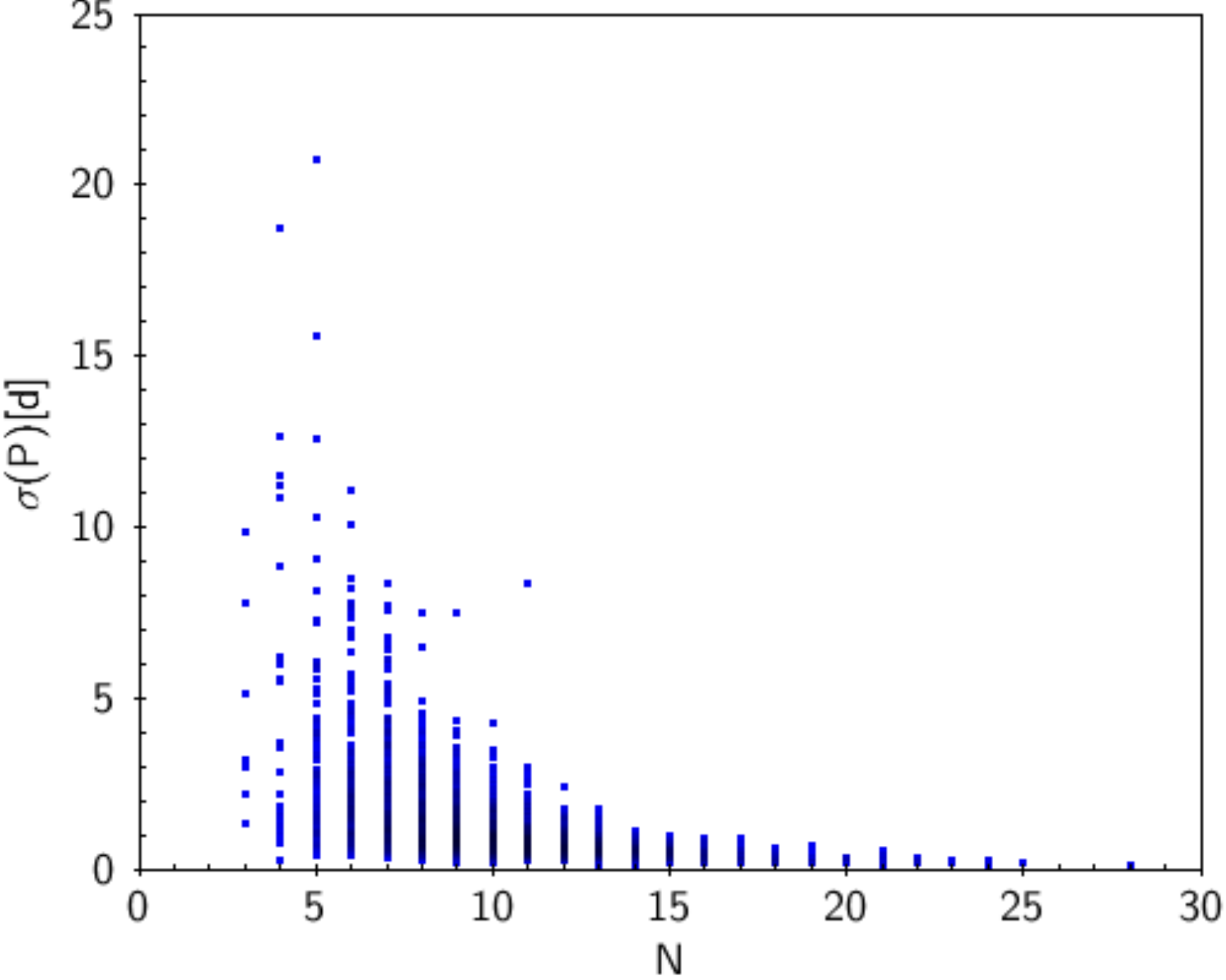} \\
\includegraphics[scale=0.32]{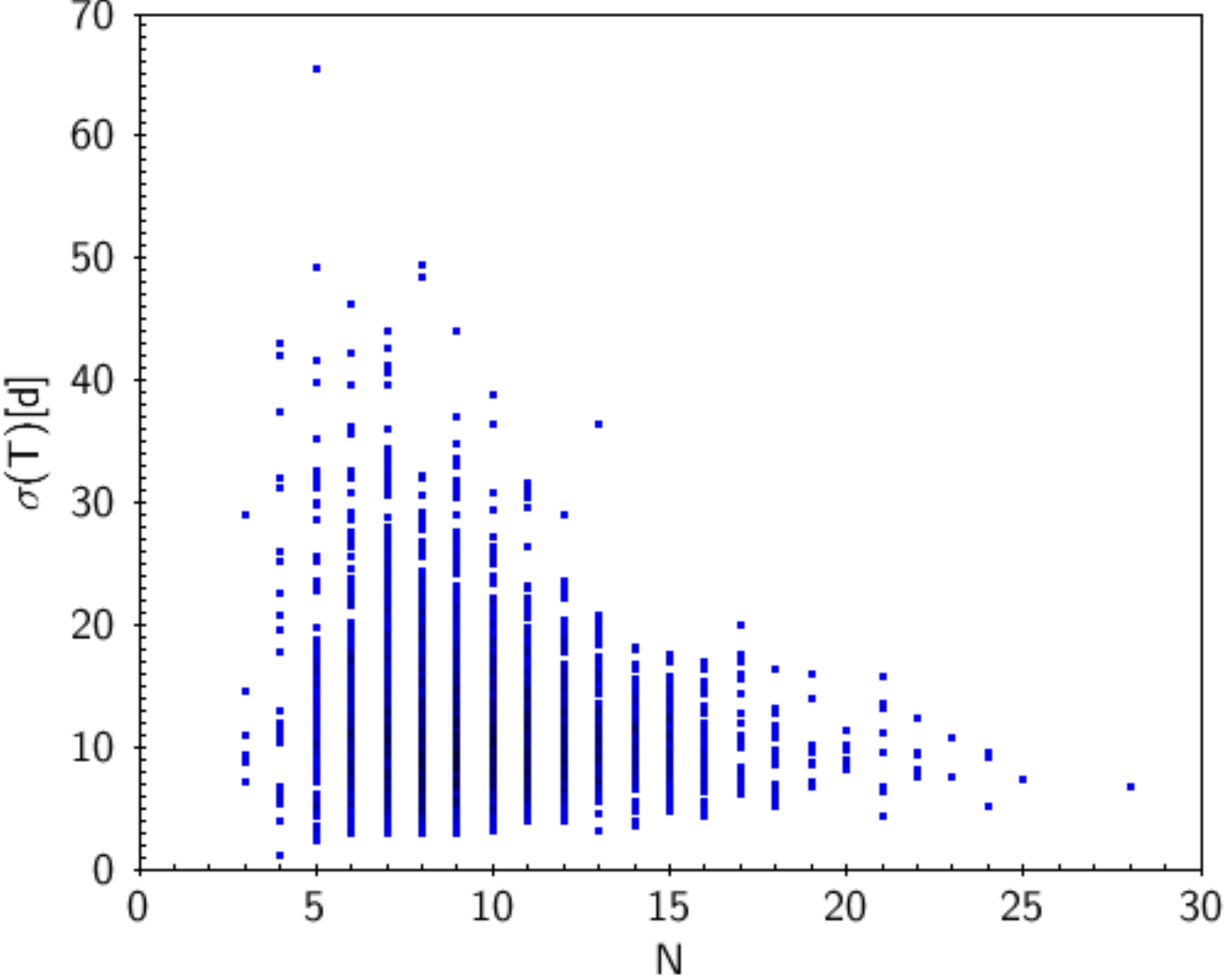} \\
\includegraphics[scale=0.32]{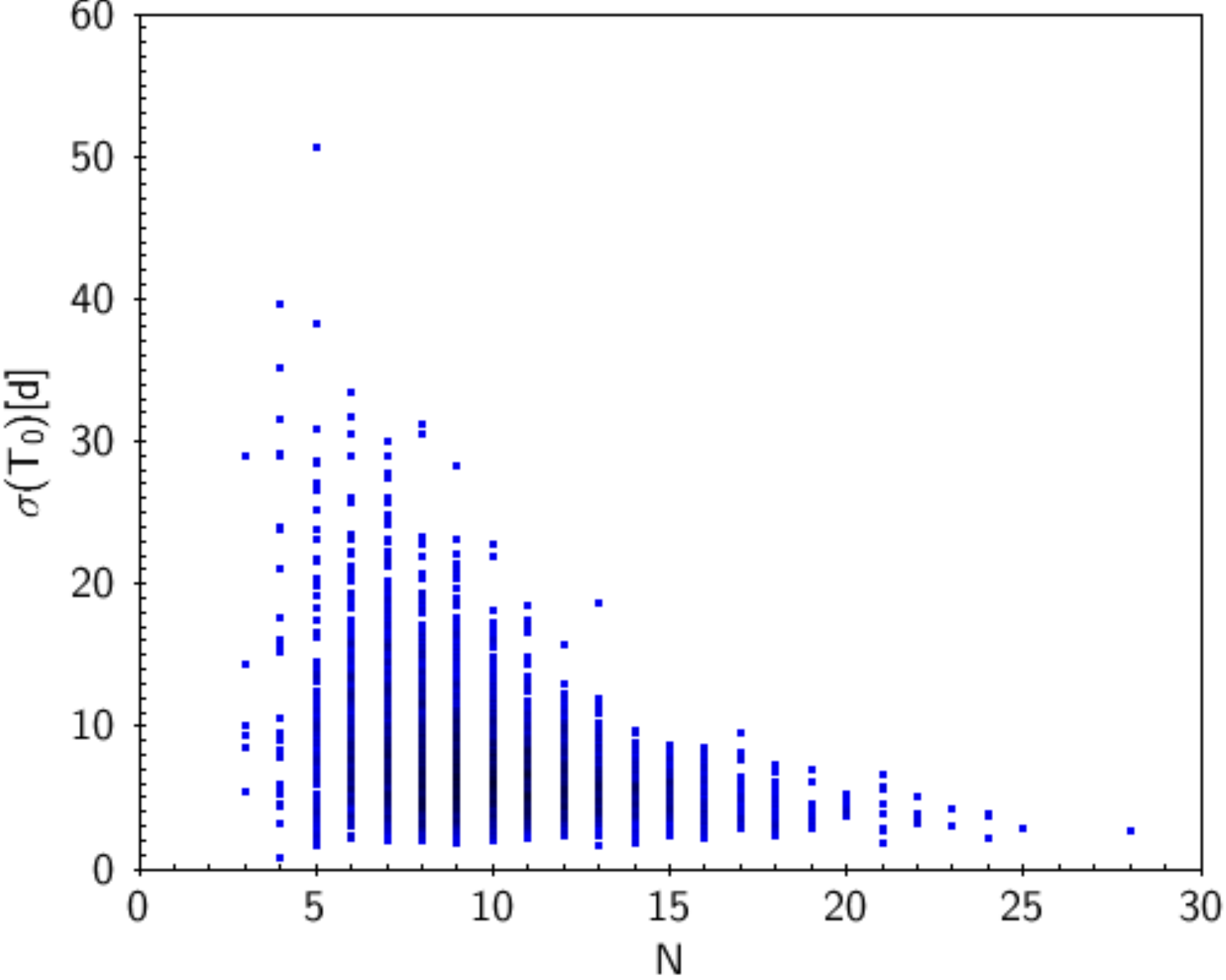} \\
\includegraphics[scale=0.32]{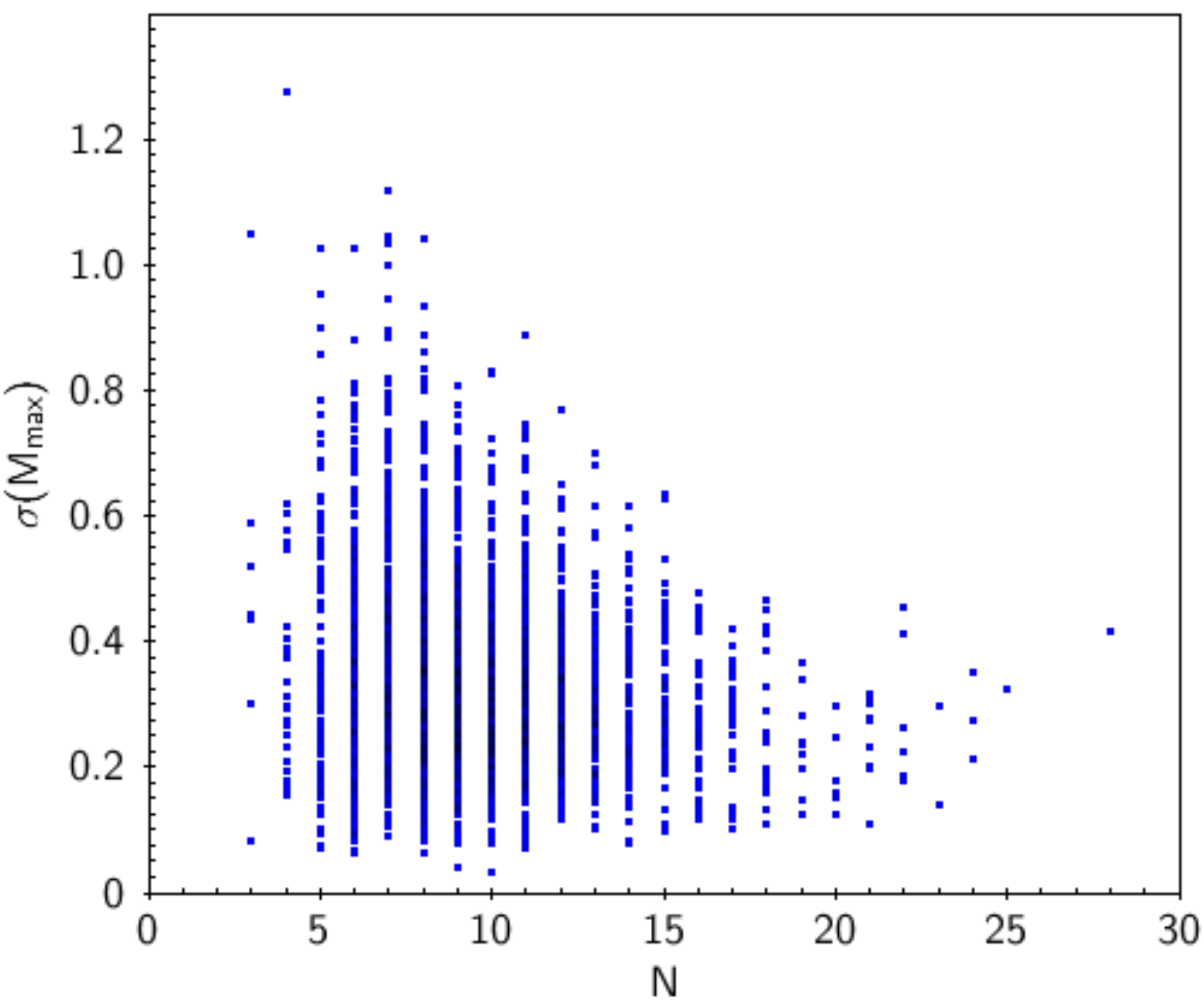} \\
\end{tabular}
\caption{Correlations between our cycle count number E and the parameters P, $\sigma$(P), $\sigma$(T), $\sigma$(T$_{0}$), and $\sigma$(M$_{\rm max}$).\label{Fig11}}
\end{figure}
 
\subsection{Internal correlations}

\begin{figure}[hbtp]
\centering
\begin{tabular}{lr}
\includegraphics[scale=0.45]{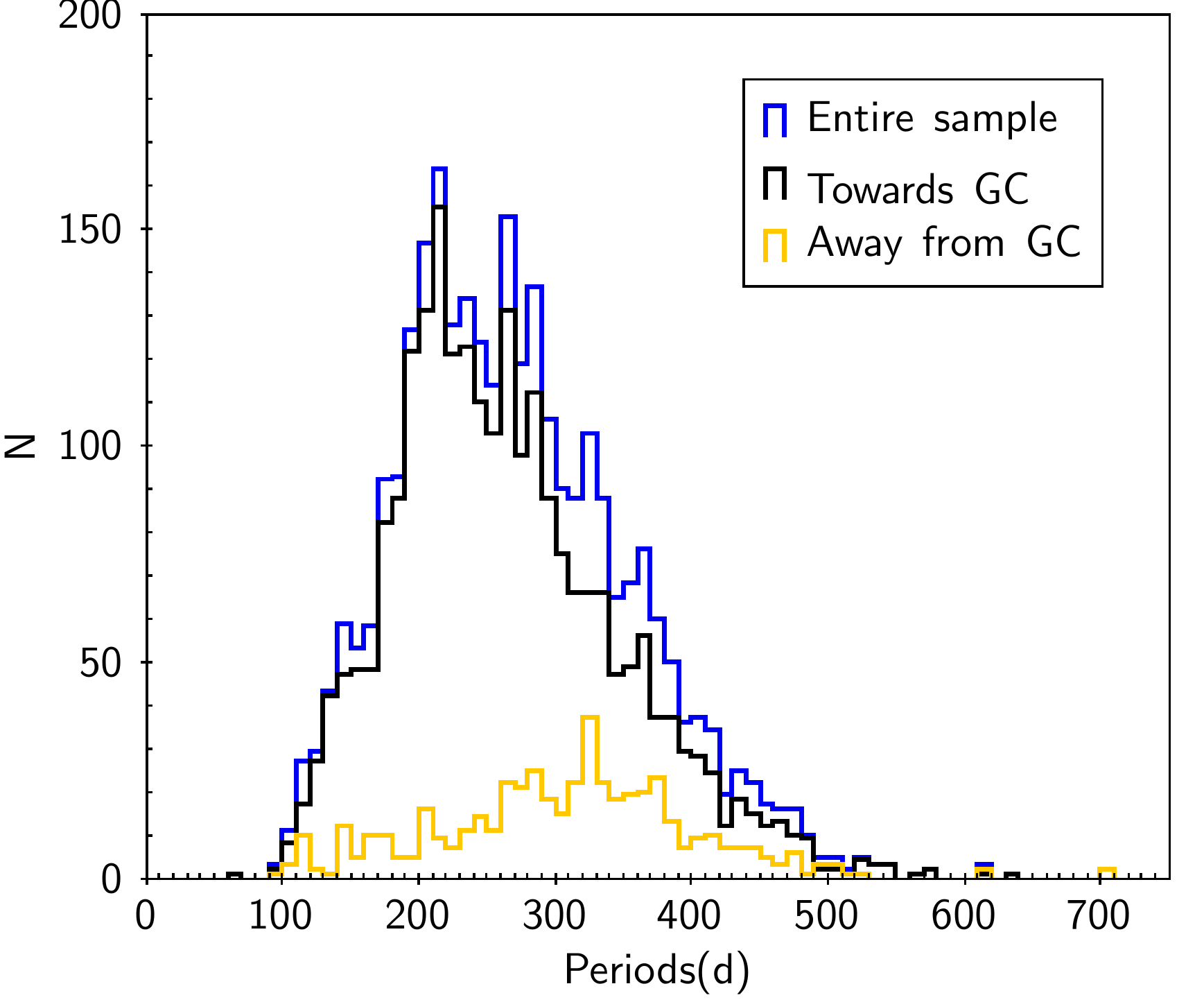} &
\end{tabular}
\caption{Period distributions towards (black) and away from (yellow) the galactic center, compared to our entire sample (blue). \label{Fig12}}
\end{figure}

In Table~\ref{table6} we give the mean values of the parameters mentioned in Table~\ref{table1}, together with their standard deviations and extremes. The correlations of period \textit{P} versus the standard deviations of \textit{P}, \textit{T} and \textit{T$_{0}$} are shown in Figure~\ref{Fig9}. Only \textit{$\sigma$(P)} seems to present a weak correlation with \textit{P}, which, however, diminishes if we consider the  ratio \textit{P} vs \textit{$\sigma$(P)/P}. Figure~\ref{Fig10} shows the relations of \textit{M$_{\rm max}$} and \textit{P} vs \textit{$\sigma$(M$_{\rm max}$)}. There seems to be a tendency of smaller standard deviations in \textit{M$_{\rm max}$} for short periods, while longer period show a large range of \textit{$\sigma$(M$_{\rm max}$)} values. Finally, in Figure~\ref{Fig11} we show correlations of \textit{P}, \textit{$\sigma$(P)}, \textit{$\sigma$(T)}, \textit{$\sigma$(T$_{0}$)} and \textit{$\sigma$(M$_{\rm max}$)} with the number \textit{N} of light maxima observed. All these diagrams show a tendency of small values for large \textit{N}, and a much higher spread of these parameters at low \textit{N}. This is mainly a selection effect, since small numbers of observations always are accompanied by larger uncertainties in the determined parameter values. The close correlation between \textit{P} and \textit{N} is an evident consequence of the fixed time interval covered by the ASAS database of about 9 years. A comparison of the our period distribution towards and away from the galactic center is shown in Figure~\ref{Fig12}, splitting our sample into those with galactic longitude \textit{l} = [270$^{\circ}$, 360$^{\circ}$], [0$^{\circ}$, 90$^{\circ}$] compared to those with \textit{l} = [90$^{\circ}$, 270$^{\circ}$]; the peak at 330 d seems to be predominant around the galactic anticenter, but it disappears at the hemisphere around the central region of the Milky Way.

\subsection{Selection of period variability candidates}

In Figures~\ref{Fig4} and~\ref{Fig5} we have marked with black dots those stars that differ from our determinations but do neither coincide with one of the other two databases, nor with one of the aliases. In the case of the VSX (Figure~\ref{Fig5}) they could be considered as potential candidates for period changes because VSX periods are, in many cases, based on observations during several decades, much longer than in ASAS. For a comparison with \citet{r12} we selected the limits according to the gaps visible in the histogram (Figure~\ref{Fig4}, upper panel). These limits are listed in the 4th column of Table~\ref{table4}. When comparing with the VSX we considered as variable candidates only those stars whose period ratios deviate more than 5\% from the nominal ones (1:2, 1:1 or 2:1). The corresponding limits are listed in the last column of Table~\ref{table4}. Table~\ref{table7} lists the 63 candidates for variability found according to these criteria. However, we should keep in mind that some of the stars with period ratio differences $<$5\% could be real variables; this is also valid for some of the cases with half or double period values in the VSX, compared to ASAS. Finally, we suppose that there should be some erroneous period values in the VSX, because our ASAS periods listed in Table~\ref{table7} have been rechecked and are rather certain. Therefore, the 63 candidates of Table~\ref{table7} require confirmation by more extended light curve data.

\subsection{A sample of target stars with multiperiodicity}

\begin{figure*}[hbtp]
\centering
\begin{tabular}{lr}
\includegraphics[scale=0.19]{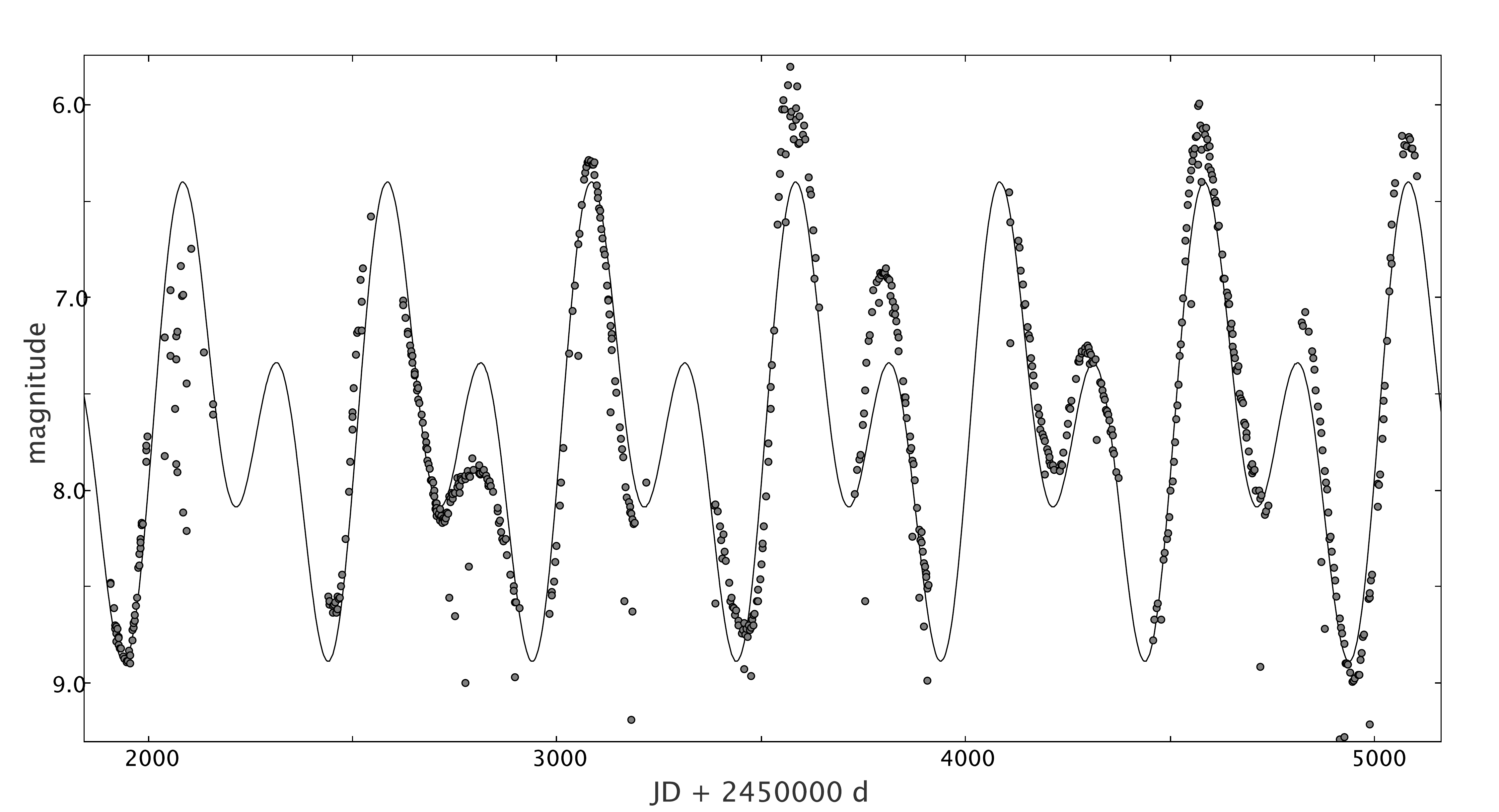} &
\includegraphics[scale=0.21]{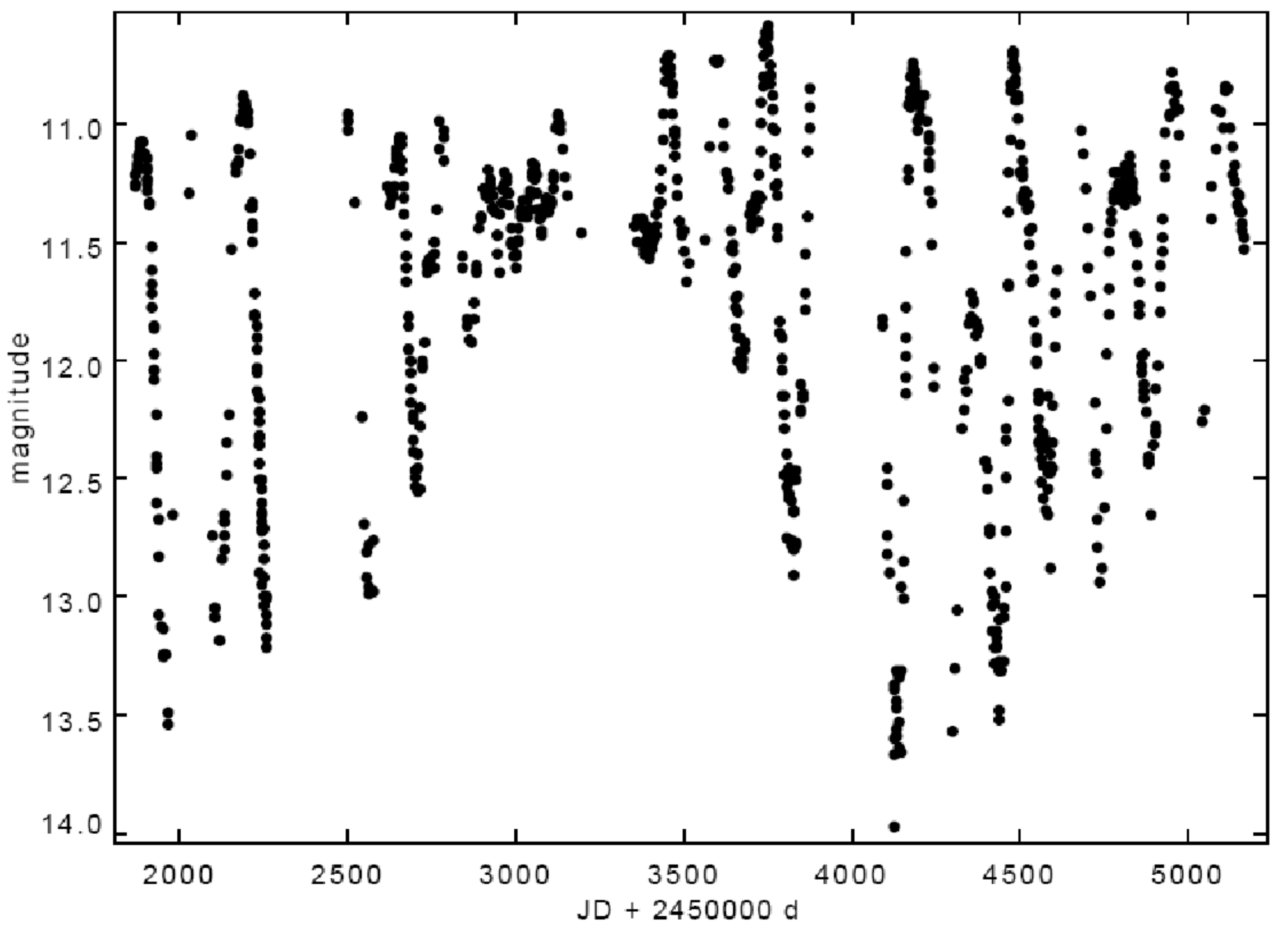} \\
\end{tabular}
\caption{\textbf{Left panel}: Light curve of R Cen showing two periods. The Fourier
spectrum reveals two peak frequencies that correspond to the periods P$_{0}$= 499$\pm$2 days and P$_{1}$=250$\pm$1 days. \textbf{Right panel}: Light curve of ASAS 054957-5252.1, showing a typical behavior of resonance due to two similar periods.\label{Fig13}}
\end{figure*}

\begin{figure*}[hbtp]
\centering
\begin{tabular}{cc}
\includegraphics[scale=0.3]{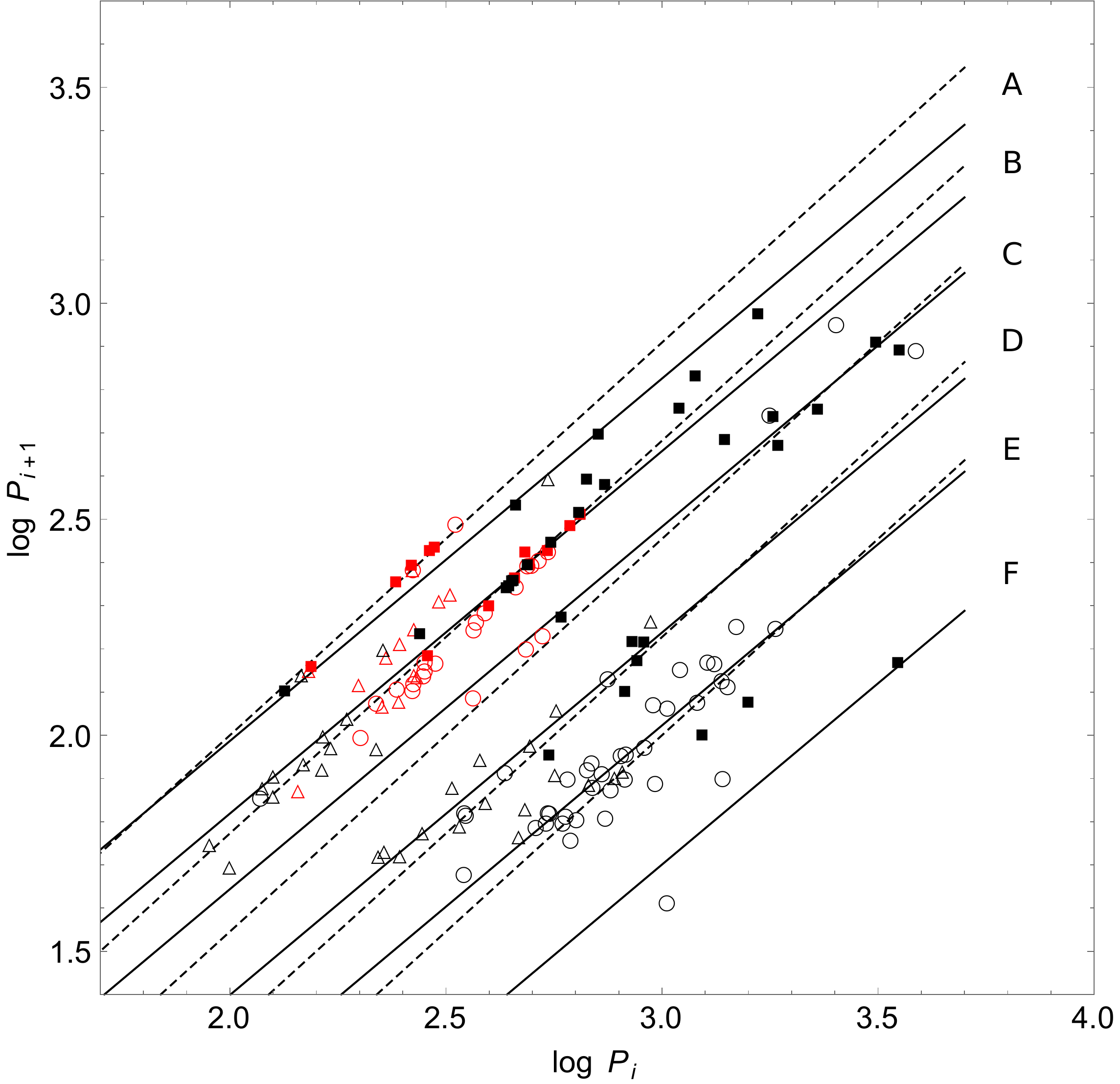} &
\includegraphics[scale=0.37]{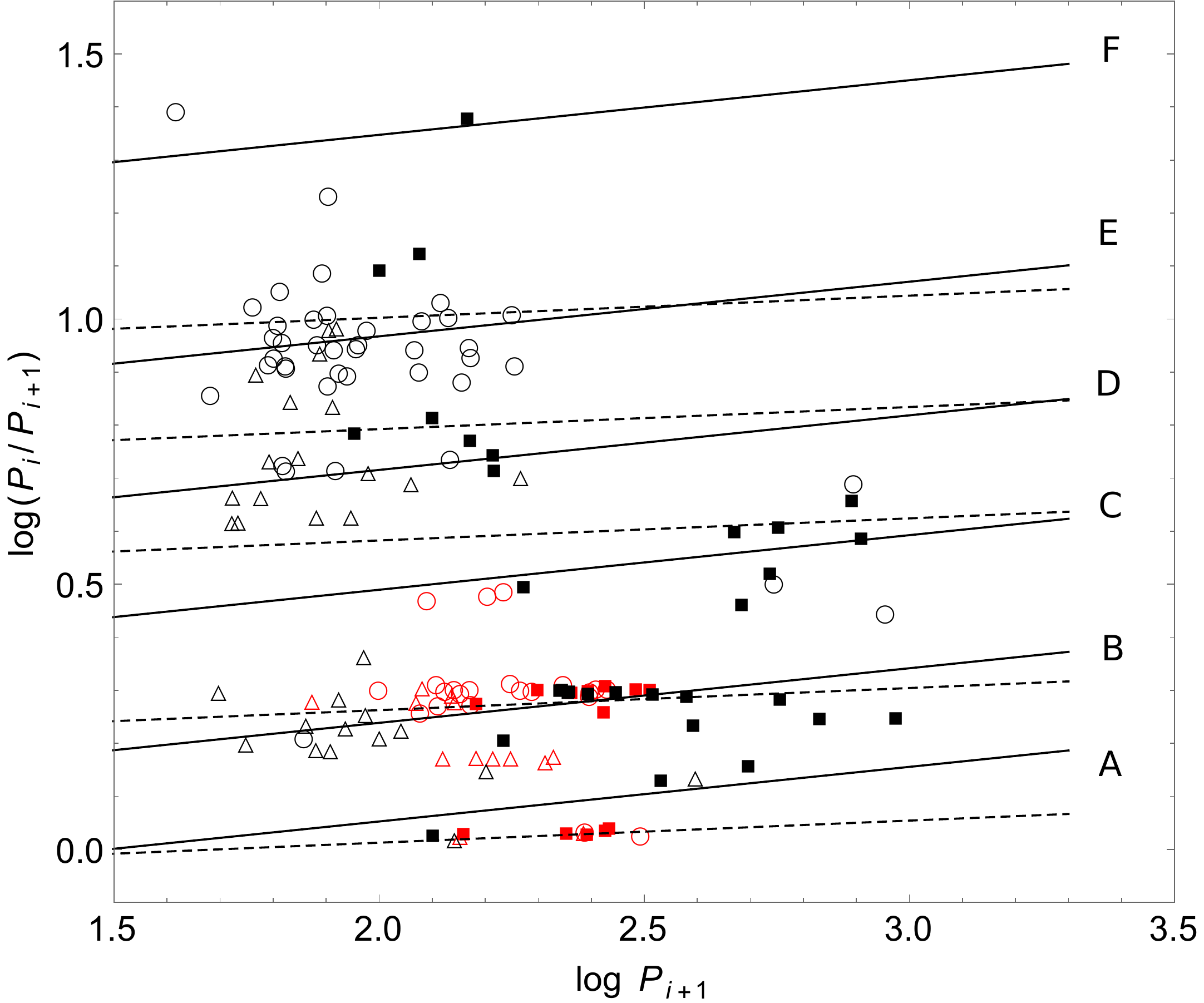} \\
\end{tabular}
\caption{\textbf{Left panel}: The Double-period diagram DPD refers to longer periods log(P$_{i}$) vs. shorter ones log(P$_{i+1}$), obtained for multiperiodic variables selected in this work (red symbols) compared with the DPD of semiregular variables found by Fuentes-Morales \& Vogt (2014, black symbols). \textbf{Right panel}: The Petersen diagram (PD) gives log(P$_{i+1}$)  vs, log(P$_{i}$/P$_{i+1}$). In both diagrams open circles refer to i = 0 for double periodic cases, filled squares to i = 0  and triangles to i = 1 for triple periodic cases. The denomination of the sequences A - F is also identical to that of Fuentes-Morales \& Vogt (2014). There are indications of an intermediate sequence between A and B for triple periodic stars.\label{Fig14}}
\end{figure*}

In the course of our ephemeris determination, we could identify multiperiodic variations for a sample of 35 stars. For each star, the dominant period was detected with our PYTHON code as described in Section 2. In order to confirm the main period and to search for additional ones we used PERIOD04 \citep{lb05}, which is based on Discrete Fourier Transform. The frequencies were detected after the main frequency was subtracted. The appearance of harmonics of main period and aliases have been controlled adjusting the calculated combined light curve to the observations. Typical examples of this analysis are shown in Figure~\ref{Fig13}. The results are listed in Table~\ref{table8}. We found 22 stars with two periods and 13 with three periods between 75 and 650 days. In order to find correlations between those multiperiods, we constructed the double period diagram (DPD) which compares contiguous periods and the Petersen Diagram (PD) which presents period ratios (Figure~\ref{Fig14}). The solid lines A to F correspond to sequences found by \citet{fv14} for SR (semi-regular) variables and dashed lines belong the sequences of \citet{k99}. The multiperiodic stars adjust to these sequences for SR stars. Apparently the sequences A and B are more populated by Mira stars compared to SR stars analyzed previously. Therefore, this sample is a valuable complement to previous investigations populating the sequences that correspond to similar periods.

\begin{deluxetable}{llllll}
\tabletypesize{\scriptsize}
\renewcommand\arraystretch{}
\tablecaption{Mean values, standard deviations ($\sigma$), and range of parameters of the 2875 stars listed in on-line Tables 1 and 2.\label{table6}}
\tablewidth{0pt}
\tablehead{
\colhead{Parameter} & \colhead{Mean} & \colhead{$\sigma$} & 
\colhead{Min} & \colhead{Max} & \colhead{Unit} \\
\colhead{} & \colhead{} & \colhead{} & 
\colhead{} & \colhead{} & \colhead{}
}
\startdata
T$_{0}$ & 2109.38 & 256.06  & 1817.01 & 3486.30 & d \\
$\sigma$(T$_{0}$) & 7.83	 & 4.51  & 0.83  & 50.71  & d \\ 
P 				  & 271.13 	 & 86.46 & 65.30& 816.74 & d \\
$\sigma$(P) 		  & 1.39 	 & 1.42  & 0.02  & 20.70  & d \\
$\sigma$(T)=O-C   & 12.47     & 6.31  & 1.12  & 65.47  & d \\
$\overline{M}_{\rm max}$  & 11.34   & 1.32  & 4.02  & 14.18  & V mag \\
$\sigma$(M$_{\rm max}$)& 0.34      & 0.15  & 0.03  & 1.28   & V mag  \\   
A 				  & 3.12      & 1.24  & 0.55  & 6.23   & V mag  \\
N 				  & 9.72      & 3.10  & 3     & 28     & - 
\enddata
\end{deluxetable}

\newpage
\section{Discussion and outlook}

When we initiated this work, we selected all 2895 stars classified as 'Mira' in the ACVS database of ASAS as our targets. Only during the analysis of the data we noticed that several stars did not fulfill the amplitude criterion for Mira classification which should be A $>$ 2.5$^{m}$ according to the definitions given in the GCVS and VSX catalogues. Apparently, many stars classified as Mira in ASAS are in fact semi-regular variables (type SR in the GCVS). However, since we could only determine amplitudes for 30\% of our targets it was impossible to reclassify the rest of our stars, and we maintained the original sample, without any attempt of variability classification. Statistically, we expect that about 20\% of all targets could be of SR type.

One of the advantages of our method is the fact that we use only the observations around the light maxima in the light curves for our analysis; they tend to have smaller errors than the data with $\sim$14$^{m}$, near the ASAS magnitude limit, as included by \citet{r12} in their analysis. This inclusion of large-scattered data could be one of the reasons of the numerous period aliases, which appeared in a considerable fraction of stars when comparing our data with \citet{r12}. Our experience also shows that, in this type of work, human interaction is still necessary; apparently, machine learning methods need further refinement, in order to achieve a quality and reliability comparable to that of generations of professional and amateur astronomers dedicated to variable star research. This latter statement is confirmed by our comparison to the VSX, which turned out to be much more reliable than that of the machine learning procedures by \citet{r12}. This, of course, refers only to the determinations of periods and amplitudes, not to the automatic variability type classification method developed by these authors and applied to ASAS. Any control or judgement of this method is outside the scope of this paper. 

\begin{deluxetable}{llll}
\tabletypesize{\scriptsize}
\renewcommand\arraystretch{0.83}
\tablecaption{Period variability candidates based on the comparison between our periods and those given in the VSX.\label{table7}}
\tablewidth{0pt}
\tablehead{
\colhead{ASAS name} & \colhead{Other name} & \colhead{P (this paper)} & \colhead{P (VSX)}
}
\startdata
024820+1730.5 		&      	T Ari  		  & 312.7 &                 	340.0 \\    
050417-2228.5 		&      	    -         & 171.8 &                 	192.0 \\
055603+1239.1 		&       QZ Ori        & 	396.07 &			        373.0 \\
063823+0813.3 		&      	    -  		  & 296.6 &     	            277.0 \\     
065138-3456.0 		&      	    -  		  & 257.5 &    	            193.0 \\   
065425-2454.2 		&      	    -  		  & 352.5 &     	            378.0 \\
071611-3753.8 		&      	SOPS IIId-19  & 222.4 &                	341.0 \\     
072838-4705.2 		&      	    -  		  & 371.7 &                 	242.0 \\     
072910-2653.3 		&      	    -  		  & 	314.6 &                 	341.0 \\
074017-4950.0 		&      	    -  		  & 284.1 &     	            308.0 \\    
074120+0822.8 		&      	U CMi  		  & 378.9 &                 	413.9 \\    
075426-4149.5 		&      	   -  		  & 322.6 &     	            349.0 \\     
080139-1501.0 		&      	   -  		  & 	448.4 &                 	472.0 \\     
080353-3836.2 		&      	 [W71b] 019-09 & 413.1 &                	520.5 \\ 
082012-3816.7 		&      	   - 		   & 526.2 &                	559.0 \\     
082426-2421.4 		&      	V561 Pup		   & 370.1 &                	270.5 \\     
082722-4201.0 		&      	   - 		   & 709.3 &                	253.0 \\     
083611-5526.7 		&      	   - 		   & 148.3 &    	            199.0  \\   
083942-2109.0 		&      	CN Pyx 		   & 258.2 &   	            300.0 \\     
084021-4449.3 		&      	BMIV89 			& 340.5 &               	316.0 \\    
084223-5443.9 		&      	   - 			& 100.1 &  	            182.9 \\     
085206-3305.7 		&      	SOPS IVd- 54 	& 369.2 &   	            256.0 \\      
090036-3456.9 		&      	    - 			& 356.9 &               	144.4 \\     
090734-2614.0 		&      	CX Pyx 			& 336.9 &               	363.0 \\ 
094717-4234.4 		&      	   - 			& 318.9 &               	335.0 \\    
102028-6618.2 		&      	   -  			& 251.1 &               	180.5 \\     
102250-4851.1 		&        [M2003b] di 1  & 382.0 &  	            405.0 \\    
103402-6118.5 		&      	[W71b] 060-11 	& 373.5 &               	146.5 \\     
104123-7628.4 		&      	Tbr V0008 		& 433.9 &  	            169.8 \\     
105148-2742.2 		&      	   - 			& 159.2 &  	            172.8 \\     
112233-4652.4 		&      	   - 			& 334.8 &  				176.6 \\     
112811-6606.3 		&      	   - 			& 350.7 &     	        389.0 \\    
113621-5843.2 		&      	   - 			& 383.9 &            	179.0 \\    
113958-5346.9 		&      	   - 			& 137.0 &                307.0 \\    
114139-5239.1 		&      	   - 			& 186.8 &  	             242.0 \\     
142611-6341.0 		&      	   - 			& 276.0 &  			 	 258.3 \\
143219-5918.8 		&      	[W71b] 095-04 	& 570.0 &                598.8 \\
151822-4951.1 		&      	   - 			& 526.1 &                590.8 \\ 
155913-2444.1 		&      	   - 			& 434.8 &                464.4 \\
160105-5021.4 		&      	Tbr V0027 		& 251.9 &    	         264.9 \\
160802-3157.3 		&      	SOPS IVd- 52 	& 387.9 &                357.2 \\
161015+2504.3 		&     	RU Her 			& 481.8 &                440.8 \\     
163255+0651.5 		&      	SS Her 			& 105.3 &    	         114.2 \\     
163355-3039.5 		&      	V910 Sco 		& 221.9 &                137.3 \\     
164821-4941.9 		&      	V630 Ara 		& 337.0 &    	         98.0  \\     
165921-1730.3 		&      	V1176 Oph 		& 222.3 &                358.0 \\
171158+0820.2 		&      	V447 Oph 		& 146.1 &    	         156.8 \\
174005-2158.9 		&      	  - 				& 326.4 &  				307.5 \\
175131-2812.0 		&      	NSV 24010 		& 451.8 &                150.2 \\     
175523-3421.0 		&      	   - 			& 202.1 &  				305.1 \\
175545-8408.5 		&      	FP Oct 			& 300.6 &                200.6 \\     
180417-0811.3 		&      	   - 			& 227.2 & 	             161.3 \\
180644-4752.3 		&      	   - 			& 201.3 &  				212.7 \\
180724-2623.6 		&      	IRC -30357 		& 381.2 &  				400.3 \\
181943+2234.9		&      	AA Her 			& 398.5 &    	        421.6 \\
182604-3010.6 		&      	   - 			& 65.3  &    	        83.8 \\     
182826-4119.9 		&      	   - 			& 382.7 &    	        357.4 \\
184632-1743.5 		&      	   -  			& 175.8 &  				116.3 \\
185331-1637.9 		&      	   - 			& 434.5 &  				477.0 \\
190716-0231.8 		&      	V1342 Aql 		& 281.6 &               	301.0   \\
193815+0331.4 		&      	MG1 1623461 		& 215.2 &               	226.8 \\
202548+0305.8 		&      	   - 			& 305.1 &  				328.3 \\
205937+2628.7 		&      	V420 Vul 		& 357.7 &               	376.5 
\enddata
\end{deluxetable}

We also found indications of  three peaks in the period distribution of targets, with maxima around 215, 275 and 330 d. According to \citet{fw14}  Mira periods dependent on the star's age or initial mass; the authors suggest mean ages of 12 Gyr at P = 200 d down to 2 Gyr at 500 d. The observed period distribution will reflect a combination of the star forming rate and the lifetimes of Mira stars of different ages. The peaks in Figure~\ref{Fig3} may indicate punctuations in the star forming history in the volume sampled. In fact, we detected that the peak at 330 d is dominating if we look away from the galactic center while it disappears in the distribution of stars towards the galactic center; here, the other two peaks at shorter periods are predominant (see Figure~\ref{Fig12}). In general, larger periods seem to be more frequent around the galactic anticenter, and shorter ones towards the center. However, a stringent statistical analysis of this item should comprise a more complete sample including stars from the northern hemisphere which is beyond the scope of this paper. We also should bear in mind that there is a 'contamination' with about 20\% semiregular variables in our sample which could be relevant in the statistics of amplitude and period distributions, as well as the multiperiodicity occurrence.

We are aware that the ASAS time span of $\sim$9 years is too short for determination of significant period changes; however, we were able, based on the comparative studies, to select candidates for period changes which are listed in Table~\ref{table7}. The 63 candidates in this Table represent about 2\% of the entire sample of 2868 stars in common with the VSX. This fraction is of the same order as some literature values:  According to \citet{zb02} about 1\% of Mira stars show evidence for period changes, but unstable periods may be more common among those with the longest periods. \citet{sz06} report also instabilities for periods $>$ 450 d and discriminate between continuous, sudden and meandering period changes. The most comprehensive study of period changes was given by \citet{t05} based in an analysis of 547 Mira stars. They found that about 10\% of them are candidates for period changes, while 1.6\% underwent strong period variability at a level $>$6$\sigma$ significance. 

These results are consistent with theoretical expectations, in favor of the idea that large period changes are being caused by thermal helium flash pulses (\citet{s13} and references therein). But there are also alternative explanations, especially for minor oscillations around a mean period, whose time scales are consistent with the Kelvin-Helmholtz cooling time of the envelope. Period changes of this sort might be caused by thermal relaxation oscillations in the stellar envelope, perhaps in response to global changes caused by a thermal pulse \citep{t05}. But accurate period values and their errors are also important for other studies, for instance in context with the mass outflow of red giants during the latest stage of their evolution. \citet{u13} found that a relation between dust mass-loss rate and Mira pulsation period exists, the largest mass loss rate values correlating with long periods. On the other hand, the mass loss rate is also related to infrared colors \citep{lw98} underlining the need of simultaneous approaches at different fronts. 

One of the most important characteristics of Mira stars is that they obey period-luminosity relations (PL); therefore, they are potentially important extragalactic distance indicators. Gaia will provide a distance calibration for all nearby Mira stars, and extensive and reliable information on their periods will be important \citep{w08,w12}. PL relations of Mira stars could provide valuable information, complementary to that given by classical Cepheids, in order to improve the distance calibration within the Milky Way, as well as to nearby galaxies.

\begin{table*}
\centering
\renewcommand\arraystretch{1}
\caption{Double and triple periodicities of selected target stars. ASAS and GCVS names, variability class according to \cite{r12}, the dominant period P$_{0}$ obtained with the PYTHON code and the longer and shorter periods $P_{\rm i}$ (${\rm i}$ = 1, 2).\label{table8}}
\begin{tabular}{cccllll}
\hline \hline 
ASAS name   & GCVS name & Variability &   P$_{\rm Mirafit}$ & $P_{0}$  &$P_{1}$  &$P_{2}$  \\ 
            &           & class       &      (d)            &(d)       &(d)      & (d)     \\                
\hline 
024745-5903.1 &X~Hor     &  Mira       & 284.9 & 283.2(0.2) & 142.5(0.2)  & \nodata     \\
035003-5721.2 &RY~Ret    &  Mira       & 128.1 & 243.2(0.6) & 226.8(0.6)  & 118.3(0.2)  \\
042757+1602.6 &W~Tau     &  Semireg\_PV & 184.6 & 243.7(0.4) & 129.3(0.5)  & \nodata     \\
054544+1838.2 &EG~Tau    &  Semireg\_PV & 219.3 & 218.4(0.5) & 119.7(0.3)  & \nodata     \\
054657+1754.5 &EI~Tau    &  Mira       & 368.2 & 366.0(0.4) & 123.2(0.9)  & \nodata     \\
054906-2241.3 & --       &  Mira       & 200.6 & 201.2(3.2) & 99.8(0.2)  & \nodata     \\
054957-5252.1 & --       &  Mira       & 131.3 & 154.9(0.1) & 144.7(0.2)  & 75.3(2.2)   \\
060401+2113.5 &V0342~Ori &  RSG        & 213.5 & 282.7(2.5) & 149.2(0.2)  & \nodata     \\
062019+1121.1 &EO~Ori    &  Semireg\_PV & 277.1 & 280.4(0.9) & 138.7(0.3)  & \nodata     \\
073753-3329.8 &KS~Pup    &  Mira       & 397.9 & 399.8(0.6) & 199.9(1.1)  & 132.7(0.6)  \\
074752-2730.0 &NSV03743  &  Mira       & 491.8 & 484.0(1.5) & 266.5(1.7)  & 244.7(10.5) \\
075751-6517.8 &X Vol     &  Mira       & 289.1 & 288.2(0.2) & 153.1(0.4)  & 143.2(0.5)  \\
082113-1229.3 &AB~Pup    &  Mira       & 267.3 & 264.9(0.2) & 128.3(20.6) & \nodata     \\
082508-4757.9 & --       &  Mira       & 486.4 & 488.9(1.1) & 249.1(12.6) & \nodata     \\
083855-4656.6 & --       &  Mira       & 311.9 & 333.0(1.0) & 311.0(0.2)  & \nodata     \\
084615-7717.4 &NSV18043  &  Mira       & 268.7 & 266.0(0.1) & 244.3(4.3)  & \nodata     \\
113359-7313.3 &NSV18797  &  Mira       & 535.3 & 545.0(1.5) & 269.1(1.3)  & \nodata     \\
113958-5346.9 & --       &  Semireg\_PV & 137.0& 298.9(1.7) & 272.9(0.3)  & 138.2(0.4)  \\
121617-5617.2 &BH~Cru    &  Mira       & 522.7 & 530.3(0.4) & 171.7(5.5)  & \nodata     \\
124902-3645.8 &V0802~Cen &  Semireg\_PV & 266.3 & 266.7(0.4) & 133.1(0.2)  & \nodata     \\
130750-6853.5 &DW~Mus    &  Mira       & 368.5 & 367.2(0.4) & 177.0(0.7)  & \nodata     \\
131935-6046.7 &TT~Cen    &  Mira       & 457.6 & 458.5(0.6) & 231.7(0.2)  & 153.6(1.5)  \\
135102-7028.4 &Z~Cir     &  Mira       & 388.8 & 389.6(0.3) & 194.2(0.5)  & \nodata      \\
141635-5954.8 &R~Cen     &  Mira       & 498.8 & 499.5(2.2) & 249.7(0.4)  & \nodata      \\
142052-6730.9 &UZ~Cir    &  Mira       & 549.2 & 546.1(0.6) & 268.5(14.3) & 178.5(3.0)   \\
151556-3935.6 &FI~Lup    &  Mira       & 265.0 & 291.6(9.1) & 268.3(0.4)  & 139.7(12.9)  \\
153557-4930.5 &R~Nor     &  Mira       & 391.6 & 495.6(0.9) & 248.8(1.1)  & 165.2(6.5)   \\
161015+2504.3 &RU~Her    &  Mira       & 481.8 & 485.2(1.4) & 160.1(1.8)  & \nodata      \\
162522-5827.8 &EQ~Nor    &  Mira       & 307.1 & 299.7(0.2) & 148.5(0.3)  & \nodata      \\
173337-3615.6 &V1163~Sco &  Mira       & 508.8 & 519.0(1.1) & 256.4(1.2)  & \nodata      \\
175412-3420.5 &BN~Sco    &  Mira       & 615.1 & 615.1(2.4) & 306.6(0.7)  & 207.1(0.6)   \\
190010-0134.9 &VX~Aql    &  Mira       & 634.7 & 650.9(2.4) & 325.4(4.3)  & 214.7(1.7)   \\
202354+0056.8 &V0865~Aql &  Mira       & 371.8 & 371.8(19.5)& 184.6(0.3)  & \nodata      \\
205300+2322.3 &RX~Vul    &  Mira       & 457.3 & 459.7(1.3) & 222.7(13.2) & \nodata      \\
222313-2203.4 &RT~Aqr    &  Mira       & 246.8 & 264.8(14.6)& 247.9(6.3)  & 121.4(8.3)   \\
\hline
\end{tabular}
\end{table*}

Most of past and present Mira star studies either refer to single or few well studied representatives, or suffer from heterogeneous data, mostly without any indication of the detailed methods and of the errors involved. We hope to deliver here a solid and homogeneous basis for future studies on periods and amplitudes of Mira stars. Those studies could extend the ASAS time interval of only 9 years towards several decades in many cases. This should be possible with modern means, as the Virtual Observatory technology involving also scanned photographic patrol plates which are available for more than one century on different places, especially at Harvard and Sonneberg observatories \citep{h99}. Another task would be the extension toward fainter magnitudes, in order to get reliable amplitudes of a much larger number of Mira stars, for instance with the Catalina Sky Survey \citep{d09,d14}, and the inclusion of other wavelengths, especially those in the infrared as done in VISTA \citep{d06,es10,s12}.

\acknowledgments
We thank our colleagues Amelia Bayo and Maja Vuckovic for valuable hints and interesting discussions. NV acknowledges the support by fund DIUV 38/2011 of Universidad de Valparaíso and by the Centro de Astrofísica de Valparaíso (CAV). We also would like to thank the referee for very useful and constructive comments that improved the presentation of the paper considerably.

\listofchanges

\end{document}